\documentclass[fleqn,usenatbib]{mnras}

\usepackage{newtxtext,newtxmath}
\usepackage[T1]{fontenc}

\usepackage{graphicx}	
\usepackage{amsmath}	
\usepackage{siunitx}
\usepackage{booktabs}

\newcommand{\aemulus}{\textsc{Aemulus}}
\newcommand{\hmass}{h^{-1}\, \mathrm{M}_\odot}

\newcommand{\hMpcInv}{{h} \mathrm{Mpc}^{-1}}

\newcommand{\derv}{\mathrm{d}}

\title[Deep learning insights into HMF]{Deep learning insights into non-universality in the halo mass function}

\author[N.-Y. Guo et al.]{
Ningyuan Guo,$^{1}$\thanks{E-mail: ningyuan.guo.20@ucl.ac.uk}
Luisa Lucie-Smith,$^{2}$
Hiranya V. Peiris,$^{3,4}$
Andrew Pontzen,$^{1}$
Davide Piras$^{5}$
\\
$^{1}$Department of Physics \& Astronomy, University College London, Gower Street, London WC1E 6BT, UK\\
$^{2}$Max-Planck-Institut für Astrophysik, Karl-Schwarzschild-Str. 1, 85748 Garching, Germany\\
$^{3}$Institute of Astronomy and Kavli Institute for Cosmology, University of Cambridge, Madingley Road Cambridge, CB3 0HA, United Kingdom\\
$^{4}$The Oskar Klein Centre for Cosmoparticle Physics, Stockholm University, AlbaNova, Stockholm, SE-106 91, Sweden\\
$^{5}$Centre Universitaire d’Informatique, Université de Genève, 7 route de Drize, 1227 Genève 4, Switzerland\\
}

\date{Accepted 2024 July 8. Received 2024 July 7; in original form 2024 May 24}

\pubyear{2024}

\usepackage{xcolor}
\usepackage{soul}

\begin{document}
\label{firstpage}
\pagerange{\pageref{firstpage}--\pageref{lastpage}}
\maketitle

\begin{abstract}
The abundance of dark matter haloes is a key cosmological probe in forthcoming galaxy surveys. The theoretical understanding of the halo mass function (HMF) is limited by our incomplete knowledge of the origin of non-universality and its cosmological parameter dependence. We present a deep learning model which compresses the linear matter power spectrum into three independent factors which are necessary and sufficient to describe the $z=0$ HMF from the state-of-the-art \aemulus{} emulator to sub-per cent accuracy in a $w$CDM$+N_\mathrm{eff}$ parameter space. Additional information about growth history does not improve the accuracy of HMF predictions if the matter power spectrum is already provided as input, because required aspects of the former can be inferred from the latter. The three factors carry information about the universal and non-universal aspects of the HMF, which we interrogate via the information-theoretic measure of mutual information. We find that non-universality is captured by recent growth history after matter-dark-energy equality and $N_{\rm eff}$ for $M\sim 10^{13} \, \mathrm{M_\odot}\, h^{-1}$ haloes, and by $\Omega_{\rm m}$ for $M\sim 10^{15} \, \mathrm{M_\odot}\, h^{-1}$. The compact representation learnt by our model can inform the design of emulator training sets to achieve high emulator accuracy with fewer simulations.

\end{abstract}




\section{Introduction}
\label{sec:introduction}

The abundance of the dark matter haloes, and by extension the abundance of groups and clusters of galaxies in the observed universe, is strongly sensitive to cosmological parameters. Accordingly, cosmological parameters can be inferred from cluster number counts \citep{Dodelson2016, Abbott2020, Abdullah2020, Costanzi2021}. To do so requires combining observational mass estimates with theoretical abundance predictions. The work of \cite{PressSchechter1974} and \cite{Bond1991} established a basic paradigm for the latter, known as the `extended Press-Schechter' (EPS) formalism. 

The underlying assumption of EPS is that dark matter haloes form at  peaks in the linear density field exceeding a pre-determined threshold. Predicting the mass of a halo is then equivalent to determining the radial extent of the corresponding density peak. 
In this picture, the effect of cosmology on halo abundance is determined by three ingredients: (i) the linear variance and therefore the abundance of peaks as a function of radius $R$, given by $\sigma(R)$; (ii) the mass $M$ enclosed within a sphere of the same radius; (iii) the height $\delta_c$ of a peak that collapses into a halo.  The first factor is fully determined by the linear matter power spectrum $P(k)$, while the second depends only on the mean matter density. Halo mass function predictions based on the first two factors without any further dependence on cosmological parameters were defined as `universal' by \citet{Jenkins2001}, and we will follow this naming convention. EPS is the prototypical example of a universal halo mass function, provided the threshold $\delta_{\rm c}$ is held fixed\footnote{Holding $\delta_{\rm c}$ fixed is justified by its weak cosmological dependence in analytic spherical collapse calculations. However, including its cosmological dependence can account for a small fraction of known non-universal behaviour \citep{Courtin2010}.}.

The original EPS formalism offers strong physical insight but very limited accuracy, with a variety of studies showing up to 50 per cent deviations between numerical simulations and predictions; see e.g. \cite{Sheth1999}, and a review in \cite{Kravtsov2012}. However, by calibrating functional freedoms on numerical simulations, far greater predictive accuracy can be achieved while retaining the essential idea of universality 
\citep{Jenkins2001, Reed2003, Tinker2008, Courtin2010, Bhattacharya2011, Watson2013, Despali2016, Diemer2020}. Universal halo mass functions generalize to new cosmological parameters, but only with an accuracy of a few per cent \citep{Courtin2010, Bhattacharya2011, McClintock2019, Bocquet2020, Diemer2020, Ondaromallea2021}. By contrast, extracting precise cosmological parameters from present and near-future survey data requires $\sim 1$ per cent accuracy on theoretical abundances \citep{Sartoris2016, McClintock2019, Euclid2023HMF}. Thus, universal halo mass functions calibrated on numerical simulations provide a good `baseline' prediction, but do not in themselves reach sufficient accuracy for cosmology requirements.

Emulators \citep[e.g.][]{McClintock2019, Nishimichi2019, Bocquet2020} offer a promising route towards the required accuracy (although other theoretical models going beyond universality, e.g. \citealt{Euclid2023HMF}, also demonstrate 1 per cent accuracy). For example, \aemulus{} (which we use in this work) reaches per cent level accuracy over much of its 7D parameter space, which extends the baseline $\Lambda$CDM parameters with the dark energy equation of state $w$ and the effective number of neutrinos $N_{\mathrm{eff}}$  \citep{McClintock2019}. However, by necessity, any emulator must place a finite number of training simulations into a cosmological parameter space which ideally would span a larger number of dimensions (to fully encapsulate the physics of dark energy, neutrinos and primordial non-Gaussianity). If the origin of non-universality were to be clearly established, the architecture and/or training of such approaches could be optimized to cover these much larger spaces with greater accuracy. 

Accordingly, a number of works have started to investigate physical contributors to non-universality (e.g. \citealt{Courtin2010, Ondaromallea2021, Euclid2023HMF, Gavas2023}), especially focusing on the potential role of the linear growth history $D(z)$ \citep{Courtin2010, Ondaromallea2021}. In the present work, we provide a complementary perspective by using an interpretable machine learning approach. Specifically, we use an \emph{interpretable variational encoder} (IVE; \citealt{Lucie-Smith2022, Lucie-Smith2023}), which is a deep learning network designed for knowledge extraction,  building on \textit{SciNet} \citep{Iten2020}. IVEs learn to compress input information into a small number of latent variables which can in turn be used to make predictions about physical quantities. For our purposes, the input is the linear matter power spectrum of a cosmological model and optionally the growth history too, and the output is the halo abundance as a function of mass. We train the IVE to reproduce HMFs at $z=0$, generated by \aemulus{} \citep{McClintock2019}, to within 0.25 per cent residuals for $M = 10^{13.2-15} \hmass$. 

The purpose of training an IVE in this way is not to supplant emulators directly, but has two linked goals. The first is to determine the  inputs and dimensionality of the latent space required to reproduce per-cent-level-accurate halo mass functions. The second is to then interpret the physical meaning of the discovered latent parameters. Specifically with respect to inputs, we are able to test whether $D(z)$ is independently informative compared to providing the IVE with $P(k)$ alone. With respect to latent interpretation, the key tool is a robust estimator of mutual information \citep[MI;][]{Piras2023}, enabling us to probe the information content of the compressed space and to distinguish non-universal from universal effects. MI is an information theoretic measure of linear and non-linear correlation between variables \citep[see e.g. a review in][]{Vergara2013}. As such, it is agnostic to any {\it a priori} expectations on what drives beyond-universal variations in the HMF within \aemulus{}{}, giving us a new perspective on this crucial question. 

This paper is organized as follows. We provide an overview of our model in Section~\ref{sec:method_overview}, in particular describing the training data and procedure in Sections~\ref{sec:training_data} and \ref{sec:IVE-training_procedure} respectively. We present the results on determining the underlying dimensionality of the latent space and the required inputs in Sections~\ref{sec:prediction} and \ref{sec:determining_inputs} respectively. We proceed with interpreting the latent representation in Section~\ref{sec:latent_representation}. We first describe the MI metric in Section~\ref{sec:MI_calculation}, and then present an overview of the information content of the latents in Section~\ref{sec:latentMI_HMF}. We focus on the latents' universal and non-universal information content in Sections~\ref{sec:latent_rep-universal_info} and \ref{sec:latent_rep-nonuniversal_info} respectively, and discuss the connection between non-universality and halo mass definition in Section~\ref{sec:mass_definition}. Finally, we discuss the results and conclude in Section~\ref{sec:discussion}.

\section{Overview of the model}
\label{sec:method_overview}

\begin{figure*}
	\includegraphics[width=0.9\linewidth]{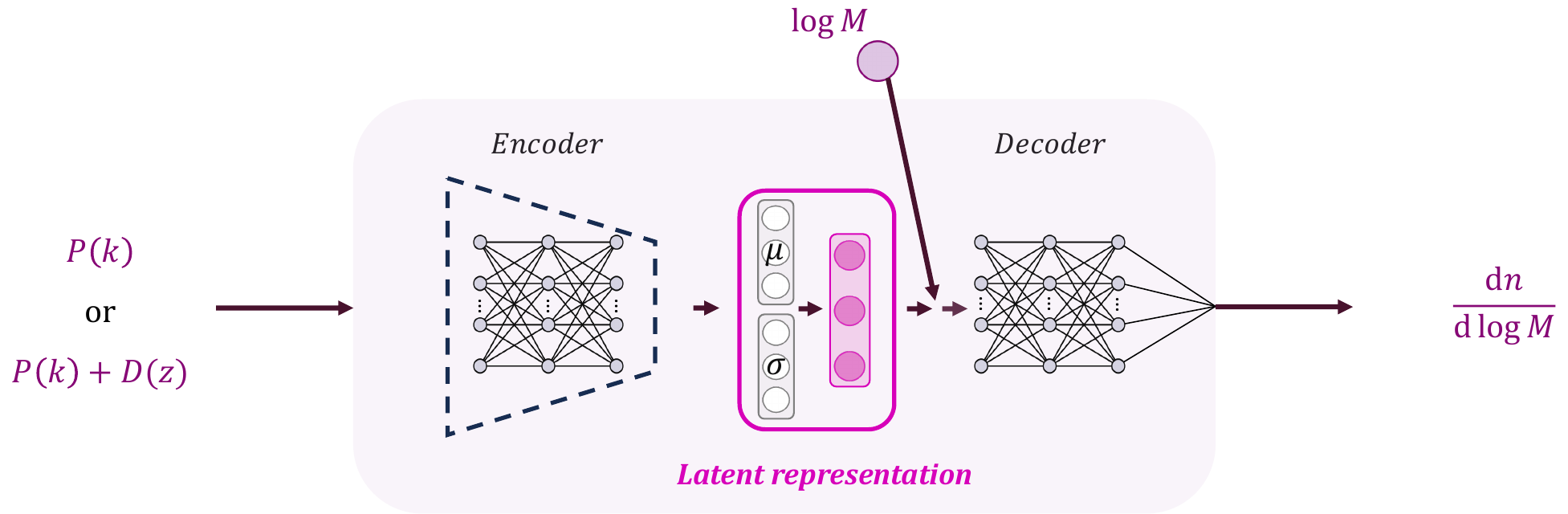}
	\caption{The interpretable variational encoder (IVE) consists of an encoder compressing the inputs (which can be the linear matter power spectrum $P(k)$ at $z=0$, or $P(k)$ concatenated with the linear growth function $D(z)$) into a low-dimensional latent representation, followed by a decoder that maps the latent representation and a given halo mass $\log M$ to the output halo number density $\derv n / \derv \log M$. The latent representation encodes all the information required to predict the HMF at $z=0$ into a set of independent latent components, which can then be interpreted in terms of their cosmological content.
 }
	\label{fig:IVE-illustration}
\end{figure*}

Our goal is to investigate the origin of the universal and non-universal information in the HMF, through the use of the IVE interpretable deep learning technique \citep{Iten2020, Lucie-Smith2022, Lucie-Smith2023}; a schematic illustration is shown in Fig.~\ref{fig:IVE-illustration}. The IVE learns to predict the $z=0$ HMF for a range of different cosmologies, when given as input e.g. the linear matter power spectrum. In doing so, it generates a low-dimensional latent representation, which captures all the information in the inputs required to predict the HMF. By investigating the cosmological information in the latents, we can gain insight into how cosmological parameters map on to the universal and non-universal aspects of the HMF. 

Since predicting the non-universal HMF requires information beyond the linear mass variance $\sigma(M)$, we provide as inputs to the IVE the linear matter power spectrum $P(k)$, and optionally the linear growth function $D(z)$ too. This choice of inputs allows us to investigate whether there is additional information about the $z=0$ HMF in $D(z)$ over information already present in the $P(k)$. 

The IVE consists of an encoder-decoder framework, similar to that in \citet{Iten2020} and \citet{Lucie-Smith2022}. The encoder is a neural network made of fully-connected layers which takes in the 1D linear matter power spectrum (or a concatenation of $P(k)$ and $D(z)$) and returns the mean and standard deviation of each Gaussian latent distribution; the decoder is another neural network made of fully-connected layers which takes in a random sample from the latent distributions and a \textit{query} halo mass $\log M$, and returns the HMF $\mathrm{d} n/\mathrm{d} \log M$. Further technical details regarding the architecture are presented in Appendix~\ref{sec:IVE-architecture}.

The IVE is similar to a variational autoencoder \citep[VAE;][]{Kingma2013}, but because its task is not to reconstruct the input, the IVE's latent space does not retain all information from the input, but only the information that is relevant to predicting the HMF. This design allows us to interpret the latent representation to extract the relevant physical factors. Relative to our previous work using IVEs \citep{Lucie-Smith2022, Lucie-Smith2023}, the model used here is simpler (using one-dimensional inputs and outputs), and is more similar to the architecture presented in \cite{Iten2020}. 
In order to interpret the latent space, we require it to be \emph{disentangled}, i.e. each latent should encode an independent factor of variation in the HMF. We achieve this by using a regularization term in the loss function (see Section~\ref{sec:loss-function}), following \citet{Iten2020} and \citet{Lucie-Smith2022}. Once the IVE model learns a disentangled latent representation, we interpret it using MI as described in the Introduction. We estimate the MI of each latent variable with the HMF, as well as with the input linear power spectrum and growth functions, to determine what information each latent variable encodes. The sub-per-cent-level accuracy requirements on the learnt mapping places stringent requirements on the accuracy of the disentangled model, which is technically challenging to achieve in comparison with these prior works; our annealing-based approach to this problem is outlined in Section~\ref{sec:IVE-training_procedure} and fully described in Appendix~\ref{app:annealing}.

\subsection{Training and test data}
\label{sec:training_data}

We now describe the construction of the ground-truth HMF outputs and the inputs used to train the IVE. We train two IVE models which differ by their inputs: the first uses only $P(k)$ as input, while the second uses both $P(k)$ and $D(z)$. Since \citet{Courtin2010, Ondaromallea2021} find information on growth history relates to non-universality, the comparison between our two models allows us to determine whether growth history is required in addition to $P(k)$ to achieve a more accurate description of the HMF. We compare the prediction accuracy of the two IVEs; if growth history provides additional relevant information, the IVE also given $D(z)$ should predict the HMF to higher accuracy.

\subsubsection{Ground truth HMFs}
\label{sec:training_data-aemulus}
We produce the HMFs to train our IVE models using the state-of-the-art HMF emulator \aemulus{} \citep{McClintock2019}. \aemulus{} can achieve 1 per cent precision over its seven-dimensional cosmological parameter space varying $\Omega_{\rm b} h^2$, $\Omega_{\rm c} h^2$, $w$, $n_{\rm s}$, $\sigma_8$, $H_0$, and $N_\mathrm{eff}$ \citep{McClintock2019}. Using \aemulus{} is particularly convenient for an exploration of HMFs with our machine learning approach, because it can generate unlimited training samples. By contrast, the largest available simulation suites only have data for a few thousand cosmological parameter samples \citep{VillaescusaNavarro2020}.  While training on \aemulus{} limits our exploration to its cosmological parameter space, this is sufficient for our  study.

To ensure we train on reliable HMFs, we choose cosmological parameter samples for our dataset to lie within the domain of validity of \aemulus{}. Following the approach that \citet{DeRose2019} used to construct the \aemulus{} training set, we project a 7D Latin hypercube into the cosmological parameter space such that the resulting samples cover approximately the $\pm 3\sigma$ region allowed by CMB+BAO+SNIa, and follow degeneracies among cosmological parameters. We provide details on the cosmological parameter sampling process in Appendix~\ref{appendix:aemulus_cosmological_parameter_space}. During training, we encountered a bug in \aemulus{} which does not affect its overall accuracy but which generates a correlation between residuals and the value of $w$; this is also described in Appendix~\ref{appendix:aemulus_cosmological_parameter_space}. 
We generate a dataset with $10^5$ cosmological parameters samples, where 48000 samples are used for the training set, 12000 samples for the validation set, and 40000 samples are used for the test set. We checked that the size of the training set we used saturated the accuracy of the learnt mapping.

The halo mass function is generally written as\footnote{Note that throughout this paper we use $\log$ to denote logarithms to base 10, and $\ln$ to denote the natural logarithm. The IVE outputs $\mathrm{d} n / \mathrm{d} \log M$, which is equivalent to equation~\eqref{eq:halo_mass_function} modulo a factor of $\ln(10)$.} \citep{PressSchechter1974, Bond1991}
\begin{equation}\label{eq:halo_mass_function}
	\frac{\mathrm{d} n}{\mathrm{d} \ln M}=\frac{\rho_{\mathrm b}}{M} \frac{\mathrm{d} \ln \sigma^{-1}}{\mathrm{d} \ln M}  f(\sigma) \,,
\end{equation}
where $n$ is the number density of haloes with mass $M$ per logarithmic mass interval, $\rho_{\rm b}$ is the background matter density of the universe, and $f(\sigma)$ is the multiplicity function. The latter gives the shape of the halo mass function in terms of $\sigma(M, z)$, the rms mass variance of the linear density field given by
\begin{equation}\label{eq:sigma_definition}
	\sigma^{2}(M, z)=\frac{1}{(2 \pi)^{3}} \int P(k, z)|\tilde{W}(\mathbf{k}, R)|^{2} \mathrm{d}^{3} k \,.
\end{equation}
Here, $P(k, z)$ is the linear matter power spectrum at the specified redshift, and $\tilde{W}$ the Fourier transform of the top hat filter with $M = \frac{4}{3}\pi\rho_{\rm b} R^3$. 

\begin{figure}
    \includegraphics[width=\columnwidth]{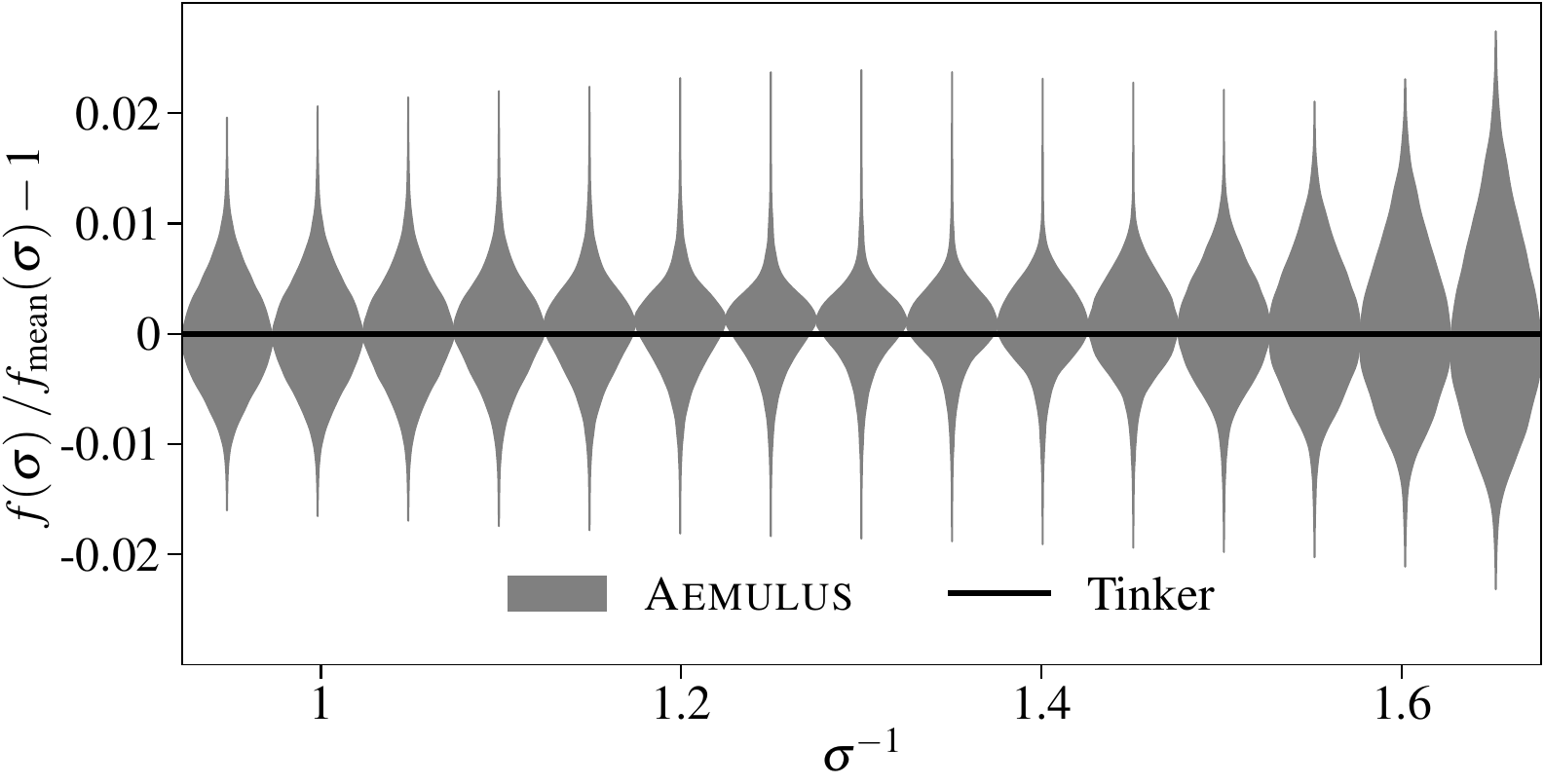}
	\caption{Distribution of $f(\sigma)$ mass functions returned by \aemulus{} (grey violin plots) and Tinker (black line) for the set of cosmological parameters used in this work, normalized by a reference mass function $f_{\rm mean}(\sigma)$ with mean cosmological parameters (see Section~\ref{sec:training_data-IVE_inputs}). Note we show \aemulus{} $f(\sigma)$ evaluated on a grid of $\sigma(M)^{-1}$ for visualisation, but the training set samples the range of halo masses and hence $\sigma(M)$ much more densely (see Section \ref{sec:training_data-IVE_outputs}). A universal HMF, such as the Tinker mass function, shows no variations in $f(\sigma)$ as a function of $\sigma(M)$. \aemulus{} captures non-universality, i.e. variations in $f(\sigma)$ at fixed $\sigma(M)$, of up to $\sim 2$ per cent for the set of cosmological parameters used in this work. 
 }
	\label{fig:ground_truth_HMFs}
\end{figure}

Given cosmological parameters, \textsc{Aemulus} outputs the HMF for haloes with masses $M_{\rm 200b}$, defined as the mass enclosed within a sphere of overdensity 200 with respect to the cosmic mean matter density. To do so, it first emulates the fitting parameters $d$, $e$, $f$, $g$ of the multiplicity function \citep{Tinker2008, Tinker2010},
\begin{equation}\label{eq:Aemulus_f_definition}
	f(\sigma)=B\left[\left(\frac{\sigma}{e}\right)^{-d}+\sigma^{-f}\right] \exp \left(-g / \sigma^{2}\right) \,,
\end{equation}
where $B$ is a normalization constant such that all dark matter resides in haloes. Then, it converts equation~\eqref{eq:Aemulus_f_definition} to the HMF via equation~\eqref{eq:halo_mass_function}, calculating $\sigma (M)$ from the linear $P(k)$. Note that \aemulus{} learns a non-universal HMF, as the fitting parameters are explicit functions of cosmology. Fig.~\ref{fig:ground_truth_HMFs} shows the distribution of $f(\sigma)$ as a function of $\sigma(M)$ for the set of cosmologies in our training sample. The effect of non-universality is at maximum of $\pm \sim 2$ per cent level. Learning this small but non-trivial effect places stringent demands on the IVE prediction accuracy; we will return to this in Section~\ref{sec:IVE-optimisation}.

\subsubsection{Inputs to the IVE}
\label{sec:training_data-IVE_inputs}
For each set of cosmological parameters in the training data, we generate the IVE inputs: the linear matter power spectrum $P(k)$ at $z=0$ and the linear growth function $D(z)$, using CAMB \citep{Lewis1999}. For modelling dark energy with $w\neq -1$, we use the Parameterized Post-Friedmann (PPF) approximation as implemented in CAMB \citep{Hu2007}.

The linear matter power spectrum $P(k)$ in units of  $h^{-3}\si{Mpc}^3$ at $z=0$ is evaluated for $10^{-4} \leq k/(\hMpcInv) \leq 10$ at 200 points logarithmically spaced in $k$. The broad range of $k$-values allows the IVE access to information at all scales, not just those that are directly encoded within $\sigma(M)$. The growth function is derived from CAMB's output using the redshift-dependent transfer function values as $D(z) = T(z, k=1 \hMpcInv)/ T(z=z_{\rm n}, k=1 \hMpcInv)$, where $z \in [0,5] $ and $z_{\rm n}$ is the redshift where the growth is normalized. We choose $z_{\rm n}=50$ to give a broad distribution of $D(z)$ at low redshifts, when growths differ the most due to dark energy. The growth function is evaluated at 100 values of $z \in [0,5]$ linearly spaced  in scale factor, which samples the lower redshifts more finely.

We perform a standard data pre-processing step to normalize our inputs and outputs to be of order unity \citep{Goodfellow-et-al-2016}.
We first calculate $P_\mathrm{mean}(k)$ and $D_\mathrm{mean}(z)$ of the mean cosmology, for which each of the seven cosmological parameters takes the mean value of its training set distribution. Specifically, the parameters of the mean cosmology are $\Omega_{\rm b} h^2 = 0.022$, $\Omega_{\rm c} h^2 = 0.118$, $w=-1$, $n_{\rm s} = 0.962$, $\ln 10^{10} A_{\rm s} = 3.087$, $H_0 = 68.3\, \mathrm{km\,s^{-1}\,Mpc^{-1}}$, $N_\mathrm{eff} = 3.44$. We then divide out the set of power spectra we obtain by $P_\mathrm{mean}(k)$, and similarly divide out the set of growth functions by $D_\mathrm{mean}(z)$. We also take the logarithm of the inputs to reduce their dynamic range. We rescale the distribution of $\log P_\mathrm{normalized}(k)$ for all $k$ in the training set to the range $[-1,1]$, and rescale the distribution of $\log D_\mathrm{normalized}(z)$ for all $z$ similarly.

\subsubsection{Outputs of the IVE}
\label{sec:training_data-IVE_outputs}
For each cosmological parameters sample in our dataset, we initially evaluate the HMF on a grid of 500 mass points linearly spaced between $\log (M/\hmass) = [13.2, 15.0]$ using the $M_{\rm 200b}$ mass definition. The mass range is chosen where \aemulus{} is most reliable: below $\log (M/\hmass) = 13.2$, the simulations that \aemulus{} was trained on did not converge to within 1 per cent with respect to simulations of higher mass resolution \citep{DeRose2019}. Above $\log (M/\hmass) =15.0$, the Poisson noise in the binned halo counts data used to train \aemulus{} is $\gtrsim 10$ per cent \citep{McClintock2019}.\footnote{The binned halo data were accessed from \url{https://github.com/tmcclintock/Aemulus_data/tree/master/aemulus_data/mass_functions}.} We use the same mass range for all cosmologies since this mass range is covered by the binned halo counts from all \textsc{Aemulus} training simulations at $z=0$. To normalize our HMFs for machine learning, we divide out the set of HMFs by that of the mean cosmology. We then take the logarithm to reduce their dynamic range.

To produce query-ground truth pairs, for the training set HMFs, we randomly sample 50 masses per HMF from the initial grid of 500 halo masses. We tested that increasing the number of queries per HMF above 50 did not significantly improve the model's prediction accuracy. For the validation set, we randomly sample 10 masses per HMF (the model training is not sensitive to this specific choice). For the test set, to evaluate the IVE predictions across the mass range, we subsample the initial grid of 500 masses by taking every 50-th mass. We additionally include the maximum mass of $M = 10^{15} \hmass$ to test the IVE performance at the upper mass bound. The distribution of $\log (\mathrm{d} n/\mathrm{d} \log M)_\mathrm{normalized}$ for all query masses in the training set is rescaled to $[-1, 1]$, and the distribution of all query masses $\log (M/\hmass)$ in the training set is also rescaled to $[-1, 1]$.

\subsection{Training procedure}
\label{sec:IVE-training_procedure}

\subsubsection{Loss function}
\label{sec:loss-function}
Training the IVE amounts to optimizing parameters of the encoder and decoder such that the model achieves high prediction accuracy and the latent representation is disentangled. This is achieved via minimizing a $\beta$-VAE loss function \citep{Higgins2017, Burgess2018}, which is written 
\begin{equation}\label{eq:loss_function}
	L =  \frac{1}{N}\sum_{N}\left(L_\mathrm{pred} + \beta \cdot \mathcal{D}_{KL}(p(\boldsymbol{\kappa}|\mathbf{x}) \| q(\boldsymbol{\kappa})) \right) \,.
\end{equation}
Here, $N$ is the number of training samples per batch and the first term $L_\mathrm{pred}$ measures the accuracy of the model's predictions, for which we adopt the mean squared error:
\begin{equation}
	 L_\mathrm{pred}  = \left(\log \frac{\mathrm{d} n_\mathrm{pred}}{\mathrm{d} \log M }- \log\frac{\mathrm{d} n_\mathrm{truth}}{\mathrm{d} \log M} \right)^{2}\,.
\end{equation}

The second term of equation~\eqref{eq:loss_function}, $\mathcal{D}_{KL}$, regularizes the latent space to encourage disentanglement.
Further details are given in Appendix \ref{app:regloss}. The balance between accuracy and regularization is controlled by the hyperparameter $\beta$, which we  tune while training to achieve both high prediction accuracy and disentanglement.

To verify that the latent variables are disentangled, we measure the MI between latent variables (see Section~\ref{sec:MI_calculation}), 
checking that it is negligible compared to the information each latent variable contains about the ground truth HMF.

\subsubsection{Optimization}
\label{sec:IVE-optimisation}
Since our goal is to discover which factors govern the HMF at $z=0$, our IVE models must be accurate enough to learn non-universality in our training data. Fig.~\ref{fig:ground_truth_HMFs} shows this is at a maximum of $\sim 2$ per cent, with a 95 per cent confidence interval of $\pm \sim 0.9$ per cent at low $\sigma^{-1}$ (low mass) and $\pm 1.3$ per cent at high $\sigma^{-1}$ (high mass). A full investigation places stringent demands on the accuracy in the HMF predictions, such that the IVE can reproduce \aemulus{} even in regimes where its accuracy is better than per cent level. Achieving this level of accuracy together with disentangled latents requires a two-stage approach involving carefully tuned annealing strategies.

The first stage aims to achieve disentanglement while improving prediction accuracy. We find models trained with a constant $\beta$ in the loss function in equation~\eqref{eq:loss_function} either do not have disentangled latent variables when the value of $\beta$ is small, or for high values of $\beta$ the latent variables are disentangled at the cost of losing prediction accuracy. We therefore use $\beta$-annealing, where we start with a high value of $\beta$ that disentangles the latents, and then slowly decrease the value of $\beta$ with each training epoch $t$ to improve prediction accuracy \citep{Shao2020Dynamic}. The slow decrease in $\beta$ allows the latents to remain disentangled while prediction accuracy improves. Further details of the annealing procedure are given in Appendix \ref{app:annealing}. Additionally, we halve the learning rate and double the batch size when the validation loss does not improve over the previous 40 epochs. We minimize the loss function in equation~\eqref{eq:loss_function} using the AMSGrad optimizer \citep{Reddi2019}.

\subsection{Determining the dimensionality of the latent space}
\label{sec:prediction}
\begin{figure}
	\includegraphics[width=\columnwidth]{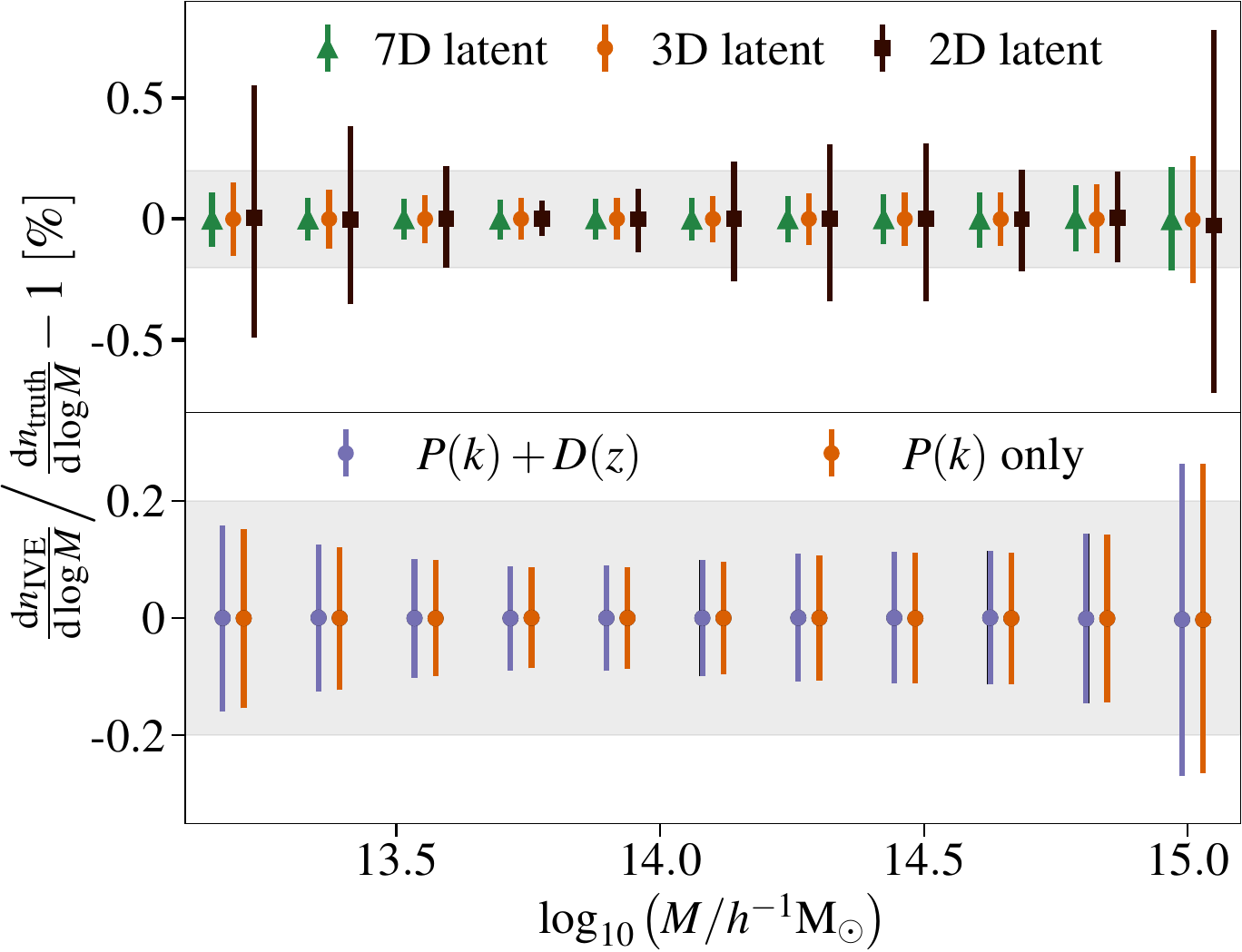}
    \caption{Mean and 95 per cent confidence interval of the HMF prediction residuals for IVE models trained using different latent dimensionalities (\textit{upper panel}), and different inputs (\textit{lower panel}). Grey shaded bands indicate $\pm 0.2$~per~cent. \textit{Upper panel:} Three latent variables can predict the HMF with similar accuracy as seven latent variables (the maximum number expected). Decreasing to two latents significantly worsens the prediction accuracy. \textit{Lower panel:} Adding $D(z)$ to the inputs does not improve the prediction accuracy compared to training on $P(k)$ alone.}
    \label{fig:prediction_accuracy}
 \end{figure}

Fig. \ref{fig:prediction_accuracy} shows the results from training the IVE using different numbers of latent variables, and for different input choices. We show the predictive accuracy of the IVE models in terms of the ratio between predicted and ground truth HMFs. The top panel shows the mean and 95 per cent confidence interval of the residuals of the predictions when training the IVE with different numbers of latent variables. We find that an IVE with a three-dimensional latent space can achieve similar accuracy to an IVE with seven latent variables; the latter is the maximum number of latent variables we expect to need, since this is the number of cosmological parameters varied to generate the training set. As we further decrease the latent space dimensionality to two, we find that the accuracy of the IVE significantly worsens. We therefore conclude that three latent variables are required and sufficient to capture all the information needed to predict \aemulus{} HMFs at $z=0$. The residuals of the three-latent-variable models are $\leq 0.25$ per cent, i.e. reaching the level of the emulator errors inherent in \aemulus{} itself. As non-universality varies HMFs in the \aemulus{} parameter range by up to $\sim 2$ per cent, to achieve this accuracy our models must have learnt information on non-universality.

\begin{figure*}
 \includegraphics[width=2\columnwidth]{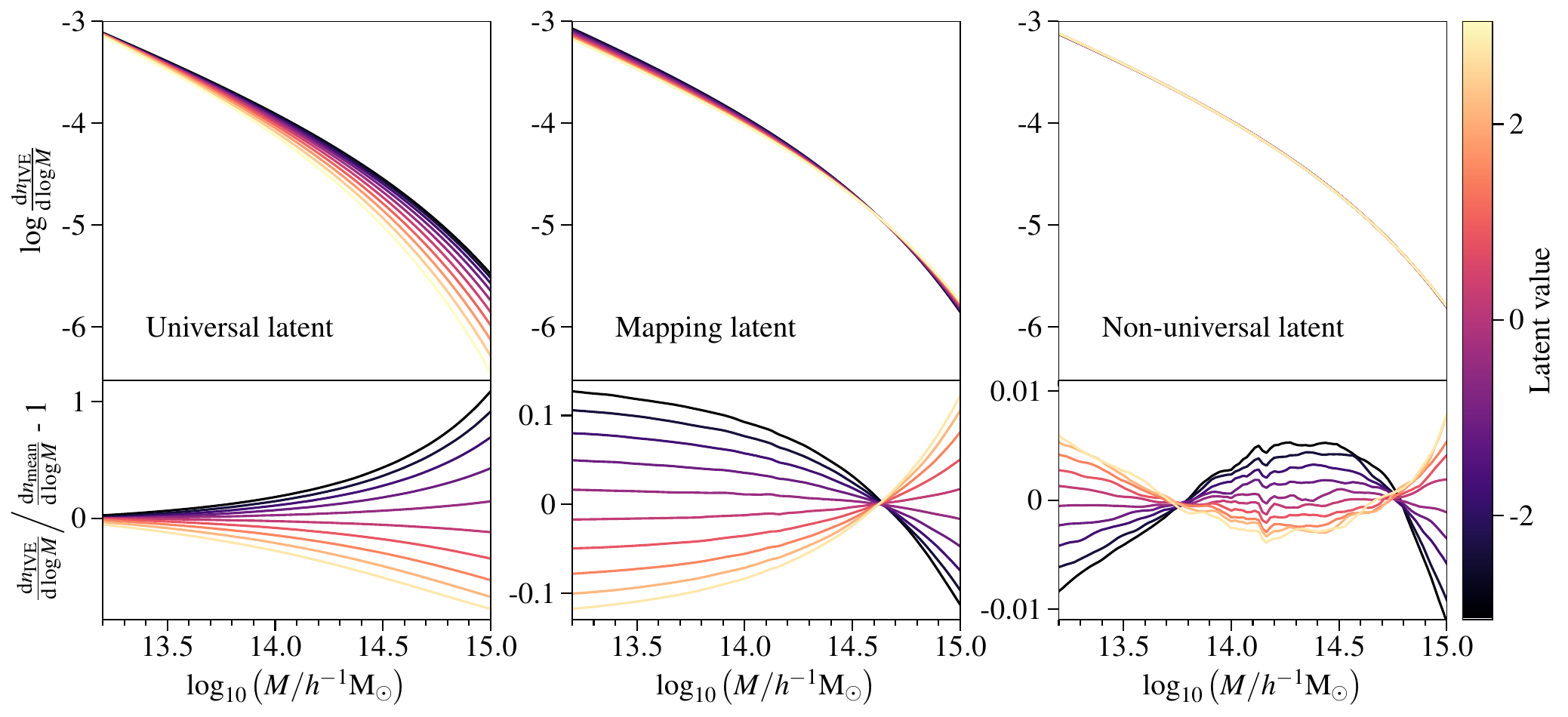}
	\caption{\textit{Upper panels:} Variations in the predicted HMF when systematically varying the value of one latent variable, while keeping the others fixed. Each panel from left to right varies the universal, mapping and non-universal latents, respectively. \textit{Lower panels:} Fractional differences in the predicted HMF with respect to that of the reference mean cosmology (defined in Section~\ref{sec:training_data-IVE_inputs}), again when varying each latent variable, keeping the others fixed. This gives an indication of a lower limit to the magnitude of each latent's effect across the parameter space as a whole.  These independent aspects of the halo mass function were discovered automatically by the IVE during training.}
	\label{fig:latent_traversals}
\end{figure*}

\subsection{Determining the required inputs}
\label{sec:determining_inputs}

The bottom panel of Fig.~\ref{fig:prediction_accuracy} compares the residuals of the two IVEs, one trained on $P(k)$ alone and the other on both $P(k)$ and $D(z)$ respectively. In both cases, three latents are necessary and sufficient to achieve the best accuracy. The IVE achieves the same accuracy regardless of whether $D(z)$ information is included in its inputs, implying that $D(z)$ does not carry additional information over that contained in $P(k)$ for describing the HMF at $z=0$ within the $w$CDM$+N_\mathrm{eff}$ parameter space considered.

This finding that $D(z)$ does not carry additional information on the HMF seems at first to contradict previous work by \cite{Courtin2010, Ondaromallea2021}, who find differences in the HMFs of simulations performed with the same power spectrum but different growth histories. However, in order to decouple the effects of growth history from the shape of the power spectrum, these studies initialized their simulations using power spectra calculated from a different set of cosmological parameters to those governing the growth history. For example, \cite{Ondaromallea2021} initialized their simulations with $P(k)$ calculated using a fixed $\Omega_{\rm m}=0.307$, while the growth history evolved according to $\Omega_{\rm m}=[0.148, 1]$. By definition, any physical effect associated with growth history is therefore unrecoverable from the power spectrum in such cases. Conversely, our approach reflects the fact that, within the \aemulus{} parameter space and where all cosmological parameters are self-consistent, it is possible to infer any relevant information directly from the power spectrum at $z=0$. 

\cite{Ondaromallea2021} further showed on a separate set of $w$CDM simulations (where $P(k)$ and growth follow consistent cosmological parameters) that including growth history in addition to $\sigma(M)$ [given in equation~\eqref{eq:sigma_definition}] improves the modelling of non-universality. However, while $\sigma(M)$ can be obtained from $P(k)$, the transformation loses information when evaluated over a finite mass range. In other words, $\sigma(M)$ for $M\geq10^{13.2} \hmass$ does not contain all the information in $P(k)$ for $10^{-4} \leq k/(\hMpcInv)\leq10$ (the relevant ranges given in Section~\ref{sec:training_data}). Hence there is no tension between previous results and our finding that $D(z)$ adds no information once $P(k)$ is known.

\section{Interpretation of learned latents}\label{sec:latent_representation}
In the previous section we presented the baseline IVE model, which takes $P(k)$ as input and encodes it within a three-dimensional latent space which is necessary and sufficient to predict \aemulus{} HMFs. We now examine the information content captured within the latent space, in order to interpret the cosmological information related to universal and non-universal aspects of the HMF.

To gain intuition on the meaning of the latent variables, we start by visualizing the impact of each latent parameter on the final HMF. To do so, we systematically vary the value of one latent, while keeping the others fixed, and show the resulting HMFs in the top panels of Fig.~\ref{fig:latent_traversals}. In each case, the chosen latent varies between $-3\sigma$ and $+3\sigma$, where $\sigma$ is the standard deviation in the latent distribution (which is close to Gaussian by construction). The plots therefore give the sought-after visual indication of the independent variation induced by each latent. However, one should bear in mind that the quantitative effect of each latent is somewhat larger than these plots illustrate, because they condition on the remaining two latents.

The panels from left to right show the impact of each latent variable in turn; the three latents are denoted `universal', `mapping' and `non-universal' respectively, for reasons that we will shortly explain. The universal latent primarily affects the amplitude of the HMF at the high-mass end; the mapping latent controls the normalization of the HMF at the low-mass end, pivoting around $M \simeq 10^{14.6}\,\hmass$; the non-universal latent controls the curvature of the HMF at intermediate mass scales. The latents exhibit a hierarchy in that the universal latent induces changes in the HMF of order 100 per cent, the mapping latent carries 10 per cent effects, while the non-universal latent carries $\sim$ 1 per cent level effects. 

In order to quantify the information carried by these latents in detail, we use tools based on the measurement of marginal and conditional MI, as we discuss below.

\subsection{Mutual information analysis}
\label{sec:MI_calculation}
We use MI both to determine whether latent variables are disentangled, and to interpret what the latent variables encode. MI is a measure of the amount of information shared between two variables $x$ and $y$, given by
\begin{equation}\label{eq:MI}
	\operatorname{MI}\left(x,y\right)=\iint  p(x, y) \ln \left[\frac{p(x, y)}{p(x)\, p(y)}\right] \mathrm{d} x \,\mathrm{d} y \,.
\end{equation}
It is zero if and only if two variables are statistically independent; see e.g. the review by \cite{Vergara2013}. Estimating MI requires approximating the joint and marginal distributions in equation~\eqref{eq:MI}. To do this, we use the GMM-MI (pronounced `Jimmie') package \citep{Piras2023}, which fits a Gaussian mixture model to the distribution of $x$ and $y$ samples. This simultaneously provides the joint and marginal densities. The package then evaluates the integral in equation~\eqref{eq:MI} using Monte Carlo integration. It also estimates the error in the MI estimate due to finite sample size by bootstrapping. The final estimate returned is the mean and standard deviation obtained from the bootstrap.

In previous work \citep{Lucie-Smith2022, Lucie-Smith2023} we demonstrated the utility of MI in the physical interpretation of IVE latent spaces. In this work, we introduce an additional tool for interpretation in the form of \emph{conditional} MI. This measures the MI between two latent variables conditioned on knowledge about additional variables; in other words, conditional MI quantifies how much information is shared amongst two variables once the conditioned variables are already known. It is defined for three variables as 
\begin{equation}
		\operatorname{MI}(x, y \mid z)=\iiint p(x, y, z) \log \frac{p(z) p(x, y, z)}{p(x, z) p(y, z)} \mathrm{d} x \, \mathrm{d} y \, \mathrm{d} z \,,
\end{equation}
where $p(x, y, z)$ is the joint probability density function of $x$, $y$ and $z$. The generalization to more than one conditioned variable follows by promoting $z$ to a vector of quantities. Generally, an increase in the MI after conditioning on $z$ means that $z$ and $x$ provide synergistic information on $y$, whereas a decrease in MI after conditioning on $z$ means that $z$ provides redundant information to $x$ about $y$. The estimation of conditional MI is also carried out using GMM-MI. Previously in astronomy, conditional MI has been used to investigate information content in multi-band galaxy surveys \citep{Chartab2023}, and to study the relation between galaxy morphology and environment \citep{Bhattacharjee2020}.

\subsection{Overview of the information content in latent space}\label{sec:latentMI_HMF}
\begin{figure*}
	\includegraphics[width=2\columnwidth]{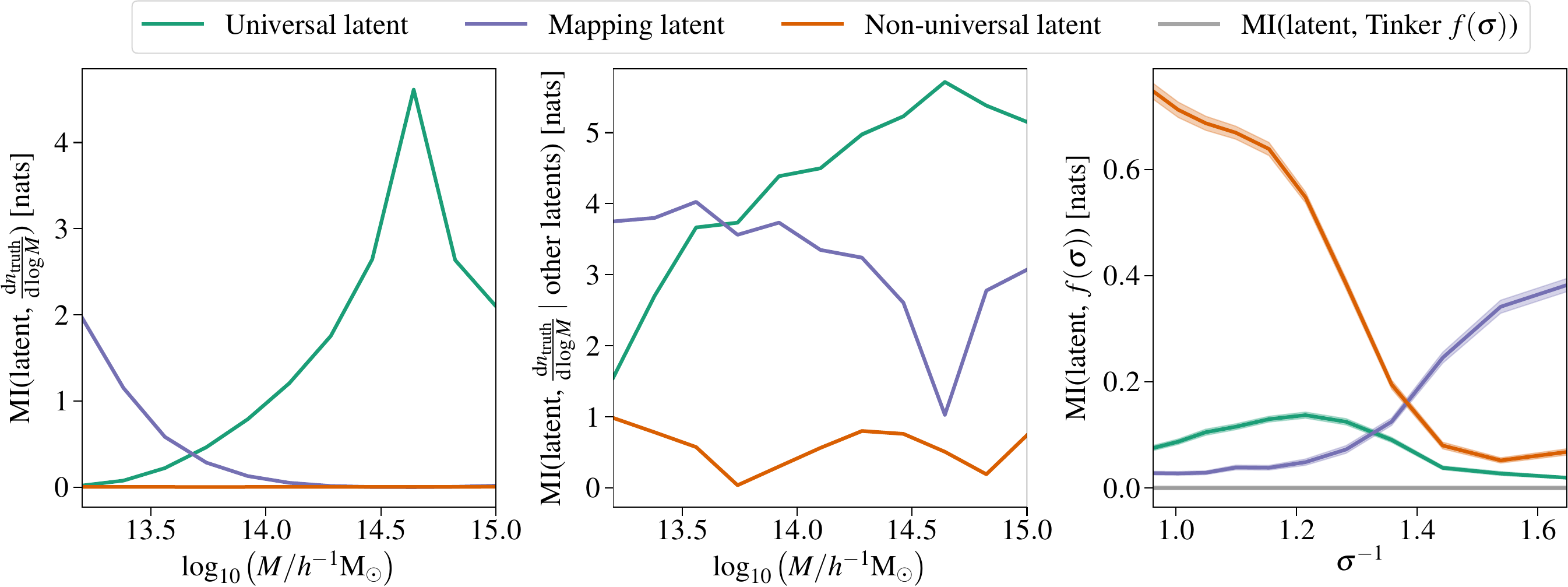} 
	\caption{\textit{Left panel:} MI between each latent variable and the ground-truth HMF in nats. Solid lines and shaded bands indicate the mean and standard deviation in the MI as estimated by GMM-MI \citep{Piras2023}. \textit{Middle panel:} MI between each latent variable and the ground-truth HMF, conditioned on the other two latent variables. \textit{Right panel:} MI between each latent and $f(\sigma)$ as function of $\sigma^{-1}$, showing the amount of non-universal information (i.e. information about variation in $f(\sigma)$ as a function of $\sigma(M)$) captured in each latent variable. The grey line shows as reference the MI between latent variables and the universal Tinker mass function's $f(\sigma)$ which, as expected, is consistent with $\operatorname{MI}=0$ and thus has no non-universal information.
	}
	\label{fig:MI_with_HMF_components}
\end{figure*}

As previously discussed in Section~\ref{sec:loss-function}, we assess the disentanglement of the latent space using MI. The models we present in Section~\ref{sec:determining_inputs} are disentangled with $\operatorname{MI} \lesssim 0.002$ nats. This disentangled 3D latent representation contains all cosmology-dependent information required to predict the (non-universal) HMF at $z=0$. The left panel of Fig.~\ref{fig:MI_with_HMF_components} shows the MI between latent variables and the ground truth $\mathrm{d} n/\mathrm{d} \log M$; the middle panel shows the MI between latent variables and the ground truth $\mathrm{d} n/\mathrm{d} \log M$ conditioned on the other two latents; the right panel shows the MI between the latents and $f(\sigma)$ as a function of $\sigma$. We now discuss the key features seen in each panel in turn.

The left panel shows that there are two dominant latents -- the universal and mapping latents -- which carry most of the information about the HMF. The nomenclature we use to describe the latents will become apparent as we analyse the latents in more detail. The universal latent predominantly carries information about the high-mass end of the HMF, in particular showing a peak at $M=10^{14.6} \hmass$ where the MI reaches $\sim 4.6$ nats.  This behaviour only emerges in  disentangled latent spaces, i.e. when we tried training with insufficient $\beta$ to obtain disentanglement, none of our latents exhibited such a clear peak. We conclude that $10^{14.6} \hmass$ is the mass-scale which best disentangles the three different factors of variation in the HMF. 

The mapping latent has an opposite trend to the universal one: its MI peaks at the low-mass end $M=10^{13.2} \hmass$ and decreases towards higher masses. The fact that the universal and mapping latents capture information about the high-mass and low-mass ends of the HMFs is consistent with the intuitive picture provided by Fig.~\ref{fig:latent_traversals}. The peak in the universal latent MI also corresponds to the pivot point in the mapping latent, so that the universal latent is almost sufficient on its own to predict the HMF at that mass. 

The information content in the third latent is highly subdominant compared to the other two. However, as we noted in the discussion of the top panel of Fig.~\ref{fig:prediction_accuracy}, this latent is necessary; the two dominant latents allow the IVE model to predict the HMF to an accuracy of $\lesssim 0.5$ per cent, but the third latent variable is required to further improve the prediction accuracy to $\leq 0.25$ per cent. We adopt the three-dimensional latent space since our goal is to reproduce results from \aemulus{} without any significant noise. The third latent is at a level somewhat below the intrinsic uncertainties in \aemulus{} itself, but we will show that it none the less encodes physically interpretable effects. 

The middle panel of Fig.~\ref{fig:MI_with_HMF_components} shows a complementary view of the information content in the latents. By conditioning the MI on the other two latents, we quantify the intuitive picture presented by  the latent traversals shown in Fig.~\ref{fig:latent_traversals}, where the effect of a single latent is illustrated by fixing the other two. Conditional MI thus reveals the effect of a single latent, which is particularly valuable in the case of the non-universal latent whose unconditioned MI is driven to zero by the dominance of the other two latents. As such, the middle panel of Fig.~\ref{fig:MI_with_HMF_components} confirms that the non-universal latent does carry additional information about the HMF. The two troughs in the non-universal latent line correspond to the dual pivot points seen in the traversal plot (right panel of Fig.~\ref{fig:latent_traversals}). Similarly, in the conditional MI view, the mapping latent contains information not only at low masses but also at high masses, with a trough corresponding to the previously identified pivot scale where the universal latent MI peaks ($M=10^{14.6}\,\hmass$). The universal latent MI retains its peak at this scale, even once conditioned on the other two latents, once again reflecting the disentanglement constraint that ensures the latents capture independent degrees of freedom.

The right panel of Fig.~\ref{fig:MI_with_HMF_components} examines the information in the latents about non-universal behaviour, by showing the MI between the latents and $f(\sigma)$. As explained in the Introduction, and following \citet{Jenkins2001}, non-universality is defined by the effect of cosmological parameters on $f(\sigma)$. Thus, any non-zero MI in the right panel of Fig.~\ref{fig:MI_with_HMF_components} indicates non-universal behaviour. Since the Tinker $z=0$ HMF is universal with no variation in $f(\sigma)$ as a function of $\sigma(M)$, the MI between the latents and the Tinker $f(\sigma)$ is consistent with zero.\footnote{Although true by construction, we have confirmed numerically that $\operatorname{MI}($latents, $f(\sigma))=0$ when using the Tinker $f(\sigma)$ using GMM-MI.} 

Strikingly, the maximal information on $f(\sigma)$ is carried by the non-universal latent, despite its subdominant effects on the overall mass function, justifying our choice of nomenclature. This is particularly true at the low-mass end (low $\sigma^{-1}$). Additional non-universal information at the high-mass end (high $ \sigma^{-1}$) is carried by the the mapping latent. On the other hand, the universal latent contains a peak of  4.6~nats of information about the HMF, but less than 0.2~nats of non-universal information; this is why this latent is referred to as the universal latent. Similarly, the mapping latent carries about 2~nats of information about the HMF peaking at the low-mass end, in comparison to the 0.4~nats of non-universal information it carries about the high-mass end. We can therefore conclude that most of the information carried by the mapping latent is universal.
We will return to a more detailed investigation of non-universality in Section~\ref{sec:latent_rep-nonuniversal_info}.

\begin{figure}
	\includegraphics[width=\columnwidth]{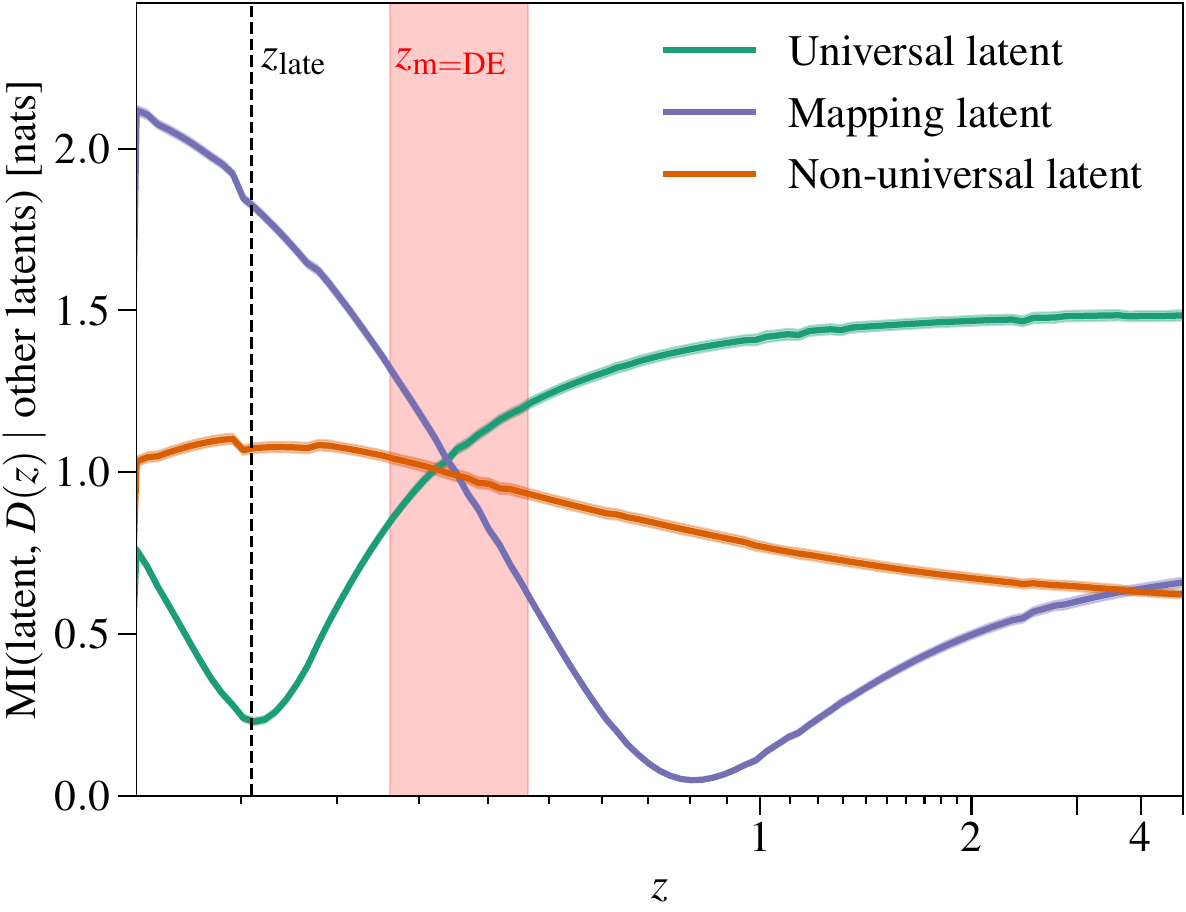}
	\caption{Conditional MI between each latent variable and the growth function $D(z)$ (normalized such that $D(0)=1$) given other latent variables. The black dashed line indicates $z_\mathrm{late} = 0.11$, and the red shaded region shows the range of the redshifts of matter-dark-energy equality $z_\mathrm{m=DE}$ in the training set.}
	\label{fig:cMI_latents_with_Dz}
\end{figure}

In Section~\ref{sec:determining_inputs} we showed that, despite being provided only $P(k)$ inputs, the latent space fully encapsulates the information about $D(z)$ required to predict the HMF. It is instructive to therefore quantify that information explicitly. 
Fig.~\ref{fig:cMI_latents_with_Dz} shows the conditional MI of each latent with $D(z)$, normalized such that $D(z=0)=1$, conditioned on the other latents.
We find that disentanglement of the latents leads to a trade-off in information before and after matter-dark-energy equality, $z_{\rm m=DE}$. This was not imposed during training, but rather discovered by the IVE when provided with only $P(k)$ information as input. This trade-off is particularly prominent for the universal and mapping latents. The non-universal latent, on the other hand, picks up information about the growth across all time.
We observe that the three latents contain similar amounts of conditional information roughly at $z_{\rm m=DE}$. Furthermore, we find that the non-universal latent's conditional MI peaks at the redshift where the universal latent's conditional MI reaches its minimum. We denote this redshift $z_{\rm late} \simeq 0.11$, and we will discuss it further in Section~\ref{sec:latent_rep-nonuniversal_info}.

\subsection{Universal information}
\label{sec:latent_rep-universal_info}
We now discuss the information content related to the universal part of the HMF that has been learnt by the latent representation, i.e. the information which could still be captured by an approximation which adopts a fixed $f(\sigma)$. As just discussed, this information is primarily encoded in the universal and the mapping latents. 

\begin{figure}
	\includegraphics[width=\columnwidth]{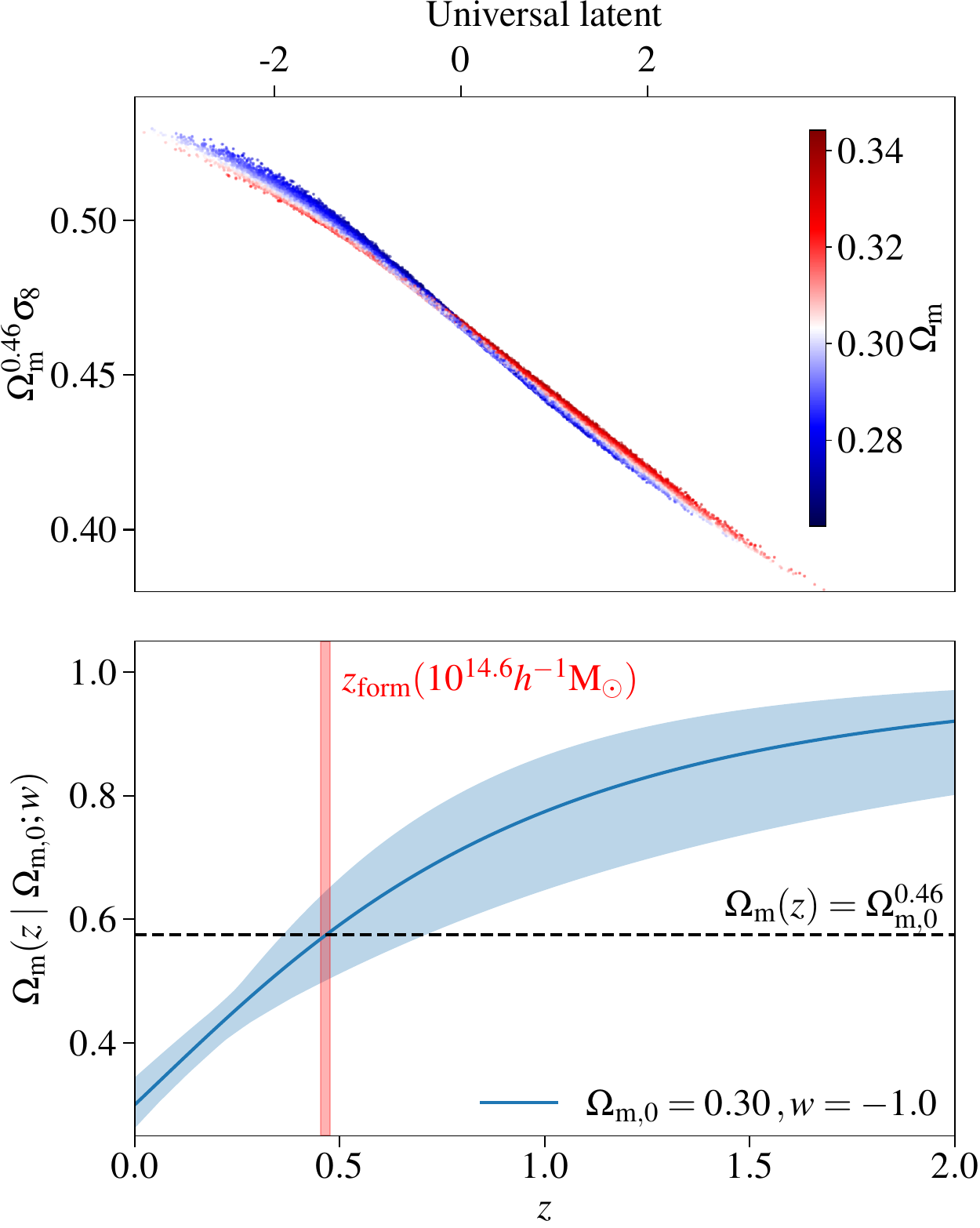}
	\caption{Cosmological parameter dependence of the `universal latent'. \textit{Upper panel:} the universal latent shows a near-perfect correlation with $\Omega_{\rm m}^{0.46} \sigma_8$; this is close to the standard $S_8 \propto \Omega_{\rm m}^{0.5} \sigma_8$ which is also known to affect the abundance of high mass haloes. The small residual scatter between the two is correlated with $\Omega_{\rm m}$. \textit{Lower panel:} $\Omega_{\rm m}(z)$ as a function of redshift $z$ for the range of cosmologies used in the training set (blue band). The red band indicates the half-mass formation redshifts $z_\mathrm{form}$ of high-mass haloes estimated from \protect\cite{Correa2015}, corrected for the $M_{\rm 200b}$ mass definition. The black dashed line indicates $\Omega_{\rm m}^{0.46}$. The fact that $\Omega_{\rm m}(z_{\rm form}) \approx \Omega_{\rm m}^{0.46}$ indicates that the latent captures the amplitude of fluctuations at the formation time of the high-mass haloes.}
	\label{fig:latentA_cosmological_dependency}
\end{figure}

We start by discussing the universal latent. The abundance of high mass galaxy clusters is known to be a sensitive probe of cosmology \citep[e.g.][]{White1993, Eke1996, Rozo2010, Norton2023}. In particular, it is common to quote sensitivity to the parameter combination $S_8 \equiv \sqrt{\Omega_{\rm m}/0.3} \sigma_8$, which encodes the amplitude of matter fluctuations within a certain redshift range \citep{JainSeljak1997}. For reasons that we will explain below, in our case the natural combination is $\Omega_{\rm m}^{0.46} \sigma_8$, which is almost equivalent to $S_8$: the exponent is nearly the same, and the $\sqrt{0.3}$ normalization in $S_8$ has no impact on information content. The top panel of  Fig.~\ref{fig:latentA_cosmological_dependency} shows a scatter plot to illustrate the close correlation between the universal latent and the parameter combination $\Omega_{\rm m}^{0.46} \sigma_8$. The scatter plot is additionally coloured by $\Omega_{\rm m}$ which captures the (small) remaining scatter in the tight relation with $\Omega_{\rm m}^{0.46} \sigma_8$.

To understand this tight relationship, recall that the MI with the HMF peaks at the mass scale $M=10^{14.6} \hmass$. Such haloes form half their mass by $z_{\mathrm{form}} \simeq 0.46$. In turn, the cosmic density at this redshift for our mean cosmology ($\Omega_{{\rm m},0}=0.30$, $w=-1.0$) is well approximated\footnote{The formation times of haloes for the different cosmologies were estimated using the analytic mass accretion history formulae in \cite{Correa2015} adapted for the $M_{\rm 200b}$ mass definition. Note that the formulae in \citet{Correa2015} were fitted on $\Lambda$CDM simulations, so strictly apply only to our mean case $w=-1$.} by $\Omega_{\rm m}(z_{\mathrm{form}}) \approx \Omega_{\rm m}^{0.46}$; this is illustrated in the lower panel of Fig.~\ref{fig:latentA_cosmological_dependency}. Using this argument, it is possible to anticipate that the parameter combination $\Omega_{\rm m}^{0.46} \sigma_8$ is an excellent proxy for information in the universal latent. 

\begin{figure}
	\includegraphics[width=\columnwidth]{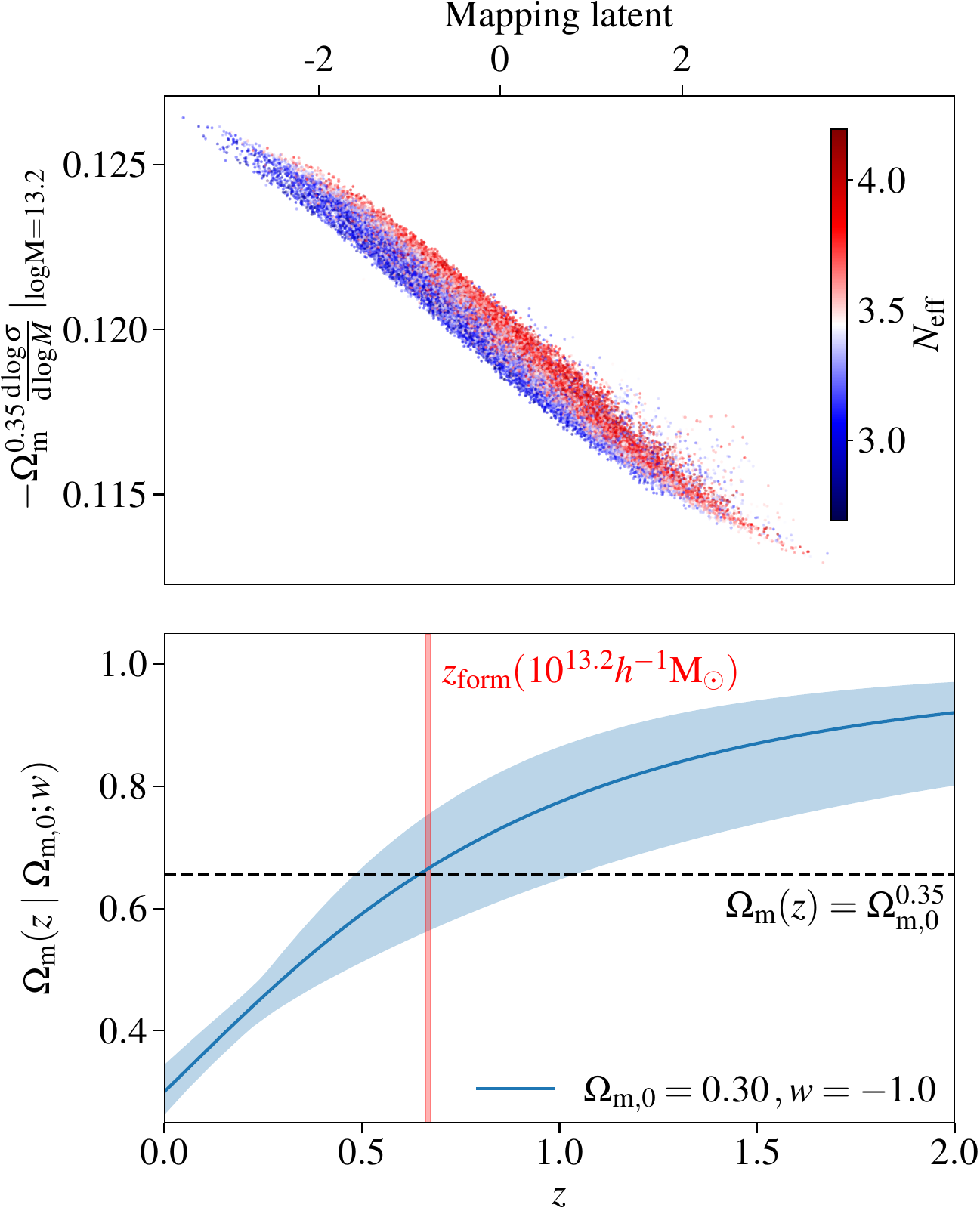}
	\caption{Cosmological parameter dependence of the `mapping latent'. \textit{Upper panel:} the mapping latent is strongly correlated with $\Omega_{\rm m}^{0.35} \frac{\mathrm{d}\log \sigma}{\mathrm{d}\log M}$ evaluated at $M = 10^{13.2} \hmass$. The residual scatter between this proxy and the latent is correlated with $N_{\rm eff}$. \textit{Lower panel:} The index of $\Omega_{\rm m}^{0.35}$ can be predicted from $\Omega_{\rm m}(z)$ of our mean cosmology (with $\Omega_{{\rm m},0}=0.3$) evaluated at the half-mass formation redshift $z_\mathrm{form}$ of $M = 10^{13.2} \hmass$ haloes. The blue band shows the range of $\Omega_{\rm m}(z)$ in the training set, and the red band indicates the half-mass formation redshifts $z_\mathrm{form}$. The black dashed line indicates $\Omega_{\rm m}^{0.35}$.}
	\label{fig:latentB-Omega_z_vs_formation_redshift}
\end{figure}

We next consider the mapping latent. Our designation for this latent derives from the hypothesis that, once the normalization is determined by the universal latent, the next most important information will concern the mapping between $f(\sigma)$ and the HMF. This hypothesis is largely borne out by our analysis, as we now explain.
First, we find that the information in this latent is well approximated by the parameter combination $\Omega_{\rm m}^{0.35}\frac{\mathrm{d}\log \sigma}{\mathrm{d}\log M}$ evaluated at $M = 10^{13.2} \hmass$. The mass scale $M = 10^{13.2} \hmass$ is where the MI between the latent and the HMF peaks, as shown in the left panel of Fig.~\ref{fig:MI_with_HMF_components}. We find that the residual scatter in the relationship between the mapping latent and this proxy is well described by $N_{\rm eff}$. This is illustrated in the top panel of Fig.~\ref{fig:latentB-Omega_z_vs_formation_redshift}.

Second, the functional form $\Omega_{\rm m}^{0.35}\, \frac{\mathrm{d}\log \sigma}{\mathrm{d}\log M}$ carries physical meaning from which the `mapping' latent nomenclature originates. The halo mass function in equation~\eqref{eq:halo_mass_function} can be written as
\begin{equation}\label{eq:halo_mass_function_for_MI}
	\frac{\mathrm{d} n}{\mathrm{d} \log M} \propto \Omega_{\rm m} \frac{\mathrm{d} \log \sigma}{\mathrm{d} \log M}  f(\sigma)\,,
\end{equation}
where the right hand side comprises of two components: the multiplicity function $f(\sigma)$ and a factor $\Omega_{\rm m} \frac{\mathrm{d} \log \sigma}{\mathrm{d} \log M}$ which maps $f(\sigma)$ to the HMF. Other than the exponent on $\Omega_{\rm m}$ (discussed below), the latter is the parameter combination found to best describe the mapping latent. Therefore, the role of the mapping latent can be thought of as a mapping between the multiplicity function $f(\sigma)$ and the halo mass function $\mathrm{d} n/ \mathrm{d}\log M$. 

Additionally, since the abundance of lower mass haloes depends more strongly on $\Omega_{\rm m} \frac{\mathrm{d}\log \sigma}{\mathrm{d}\log M}$ than on $f(\sigma)$, this result is also consistent with the finding that the mapping latent carries mostly information on the abundance of lower mass haloes (see left panel of Fig.~\ref{fig:MI_with_HMF_components}). Similar to the universal latent case, we find that the exponent of 0.35 in $\Omega_{\rm m}$ can be predicted from the formation history of low-mass haloes as shown in Fig.~\ref{fig:latentB-Omega_z_vs_formation_redshift}: at the redshift $z_{\rm form}$ where haloes of $M=10^{13.2} \hmass$ form half their mass, $\Omega_{\rm m}(z_{\rm form}) \approx \Omega_{\rm m}^{0.35}$ for the mean $\Omega_{\rm m}=0.3$ in the training set. 

In principle, one could extend the $\Omega_{\rm m}^n \frac{\mathrm{d}\log \sigma}{\mathrm{d}\log M}$ parameter combination to also include a dependence on $N_{\rm eff}$ either by manual search or automated symbolic regression techniques, in order to achieve a near-perfect linear correlation with the mapping latent (as found for the universal latent). We choose not to pursue this route in order to restrict ourselves to describing the latents in terms of physically interpretable parameter combinations only, rather than empirically motivated parameter combinations.

\begin{figure*}
    \includegraphics[width=2\columnwidth]{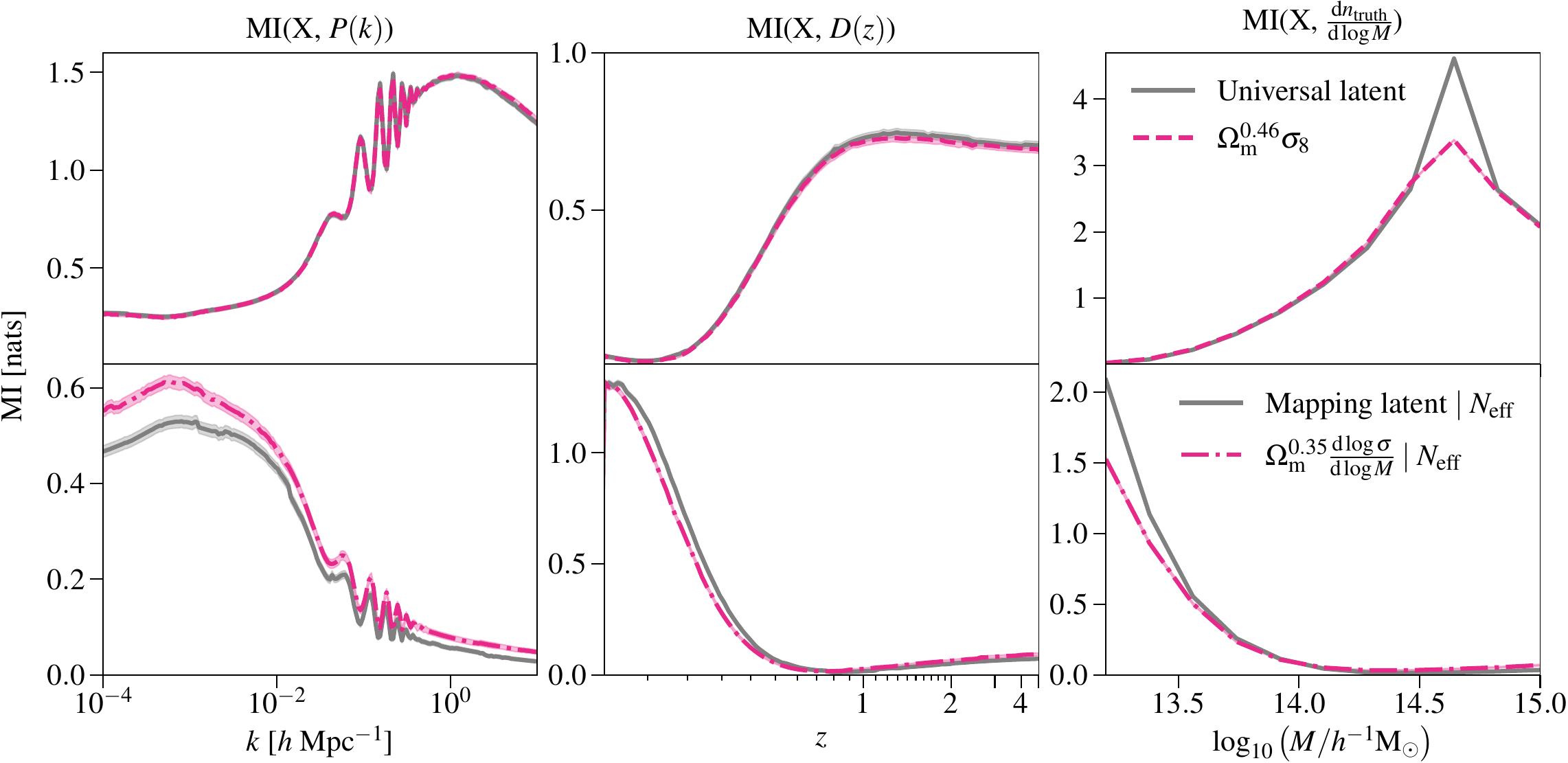}
	\caption{\textit{Top row:} MI between the universal latent and three different functions: the linear matter power spectrum $P(k)$ (\textit{left}), the growth function $D(z)$ normalized to unity at $z=0$ (\textit{middle}), and the ground truth HMF (\textit{right}). In pink dashed line, we show the same MI quantities but using $\Omega_{\rm m}^{0.46} \sigma_8$ instead of the universal latent. \textit{Bottom row:} MI between the mapping latent and the same three functions, conditioned on $N_\mathrm{eff}$. In pink dash-dotted line, we show the same conditional MI quantities but using $\Omega_{\rm m}^{0.35}\frac{\mathrm{d}\log \sigma}{\mathrm{d}\log M}$ evaluated at $M = 10^{13.2} \hmass$ instead of the mapping latent.}
	\label{fig:MI_latentAB_vs_approx}
\end{figure*}

In Fig.~\ref{fig:MI_latentAB_vs_approx}, we additionally use MI to show the close resemblance between the parameter dependencies identified and the main information content in the latents.

In the top row of Fig.~\ref{fig:MI_latentAB_vs_approx}, we compare MI between the universal latent and $P(k)$, $D(z)$ and the HMF with that of its main parameter dependence, $\Omega_{\rm m}^{0.46}\sigma_8$. We find that the MI curves of the latent and its proxy cosmological parameter dependence overlap closely. This agrees with the tight correlation seen between $\Omega_{\rm m}^{0.46}\sigma_8$ and the universal latent in the top panel of Fig.~\ref{fig:latentA_cosmological_dependency}, even without accounting for the extra small dependence on $\Omega_{\rm m}$.

The bottom row of Fig.~\ref{fig:MI_latentAB_vs_approx} compares the MI between the mapping latent and $P(k)$, $D(z)$ and the HMF, with that of its proxy, $\Omega_{\rm m}^{0.35}\frac{\derv \log \sigma}{\derv \log M}$ evaluated at $M = 10^{13.2} \hmass$. We account for the mapping latent's additional dependence on $N_{\rm eff}$ by conditioning on it. We find that the two curves are very similar, confirming that this parameter combination is a good proxy of the cosmological information in the mapping latent.

We note that the mapping latent, which has the highest MI with the HMF at the low-mass end, encodes high information on $P(k)$ at small $k$ (large scales). In contrast, the universal latent, which encodes the most information on the HMF at the high-mass end, has high information with $P(k)$ at larger $k$ (smaller scales). This may seem counter-intuitive at first, but it can be explained by the main parameter dependence of each latent. The universal latent has high MI with $\sigma_8$; from equation~\eqref{eq:sigma_definition}, we expect both very large scales and small scales to contribute minimally to the integrand; this is indeed the case in the top left panel of Fig.~\ref{fig:MI_latentAB_vs_approx}. On the other hand, the mapping latent has high MI with $\Omega_{\rm m}$. The power spectrum is affected on all scales by $\Omega_{\rm m}$, but on small scales it is additionally affected by e.g. the baryon density and $N_{\rm eff}$. Hence, $P(k)$ is most informative of $\Omega_{\rm m}$ on large scales, leading to a high MI between the mapping latent and $P(k)$ on those scales. This also leads to the high MI between the mapping latent and $D(z)$ at low redshifts, when the growth function normalized at $z=0$ is most strongly dependent on $\Omega_{\rm m}$.

\subsection{Non-universal information}
\label{sec:latent_rep-nonuniversal_info}

\begin{figure}  
 \includegraphics[width=\columnwidth]{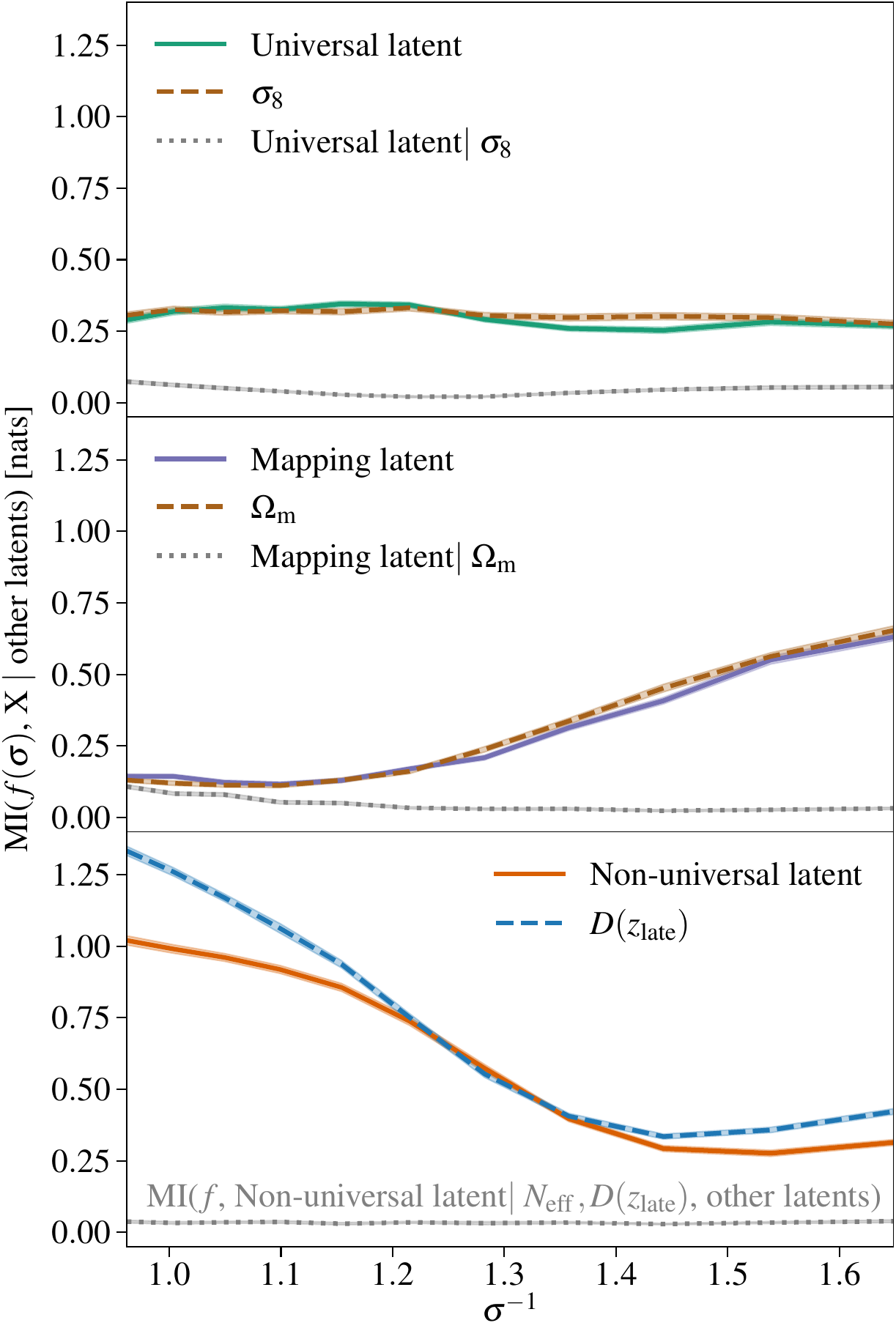}
	\caption{Conditional MIs between $f(\sigma)$ and various quantities, for interpreting the non-universal information content of the latents. The MIs are shown as functions of $\sigma^{-1}$, which increases with mass. From top to bottom, the panels show the universal, mapping and non-universal latents respectively, in each case conditioned on the remaining two latents. The leading order dependencies for the latents are encoded respectively by $\sigma_8$, $\Omega_{\rm m}$ and $D(z_{\rm late})$, as illustrated by the dashed lines. By further conditioning the MI between the latent and $f(\sigma)$ on  these leading-order dependencies, we obtain the grey dotted lines, demonstrating that there is then little information remaining. In the case of the non-universal latent (bottom panel), we need to condition on $N_{\mathrm{eff}}$ as well as $D(z_{\rm late})$ to reach this point of information saturation.
 }
	\label{fig:non-universality_in_dominant_latents}
\end{figure}

In the previous section, we discussed the `universal' information in the first two latents. However, as discussed in Section~\ref{sec:latentMI_HMF}, the latent variables also capture information on non-universality, i.e. information that is not present in $\sigma(M)$. This was already demonstrated in the right panel of Fig.~\ref{fig:MI_with_HMF_components} through the MI between each latent variable and $f(\sigma)$. The plot further demonstrates that there is an inversion in the hierarchy of information: the universal latent which dominates the overall information content of the halo mass function learns the least out of all three latents about non-universality. 

We now consider the interpretation of the non-universal information in the latents. Fig.~\ref{fig:non-universality_in_dominant_latents} shows the MI between each latent and $f(\sigma)$, but now conditioned on the other two latents to isolate the effects. The panels from top to bottom illustrate the universal, mapping and non-universal latents respectively. 

The top panel reconfirms that the universal latent carries little information about non-universality (solid line), since the MI between the latent and $f(\sigma)$ is approximately 0.25 nats for all $\sigma^{-1}$ values. We previously showed that the universal latent carries almost the same information as the parameter combination $\Omega_{\mathrm{m}}^{0.46} \sigma_8$. When examining its contribution to non-universality, we find that the information content closely mimics that in $\sigma_8$ (dashed line). This is further demonstrated by calculating the MI between the latent and $f(\sigma)$ but conditioning on $\sigma_8$ (in addition to the other latents). The grey line indicates that this conditional MI is close to zero at all scales, meaning that there is negligible further information about non-universality in the universal latent once $\sigma_8$ is known. However, since the overall contribution to non-universality is anyway concentrated in the other two latents, the overall effect of $\sigma_8$ on non-universality will be strongly subdominant and we do not consider it further. 

We next consider the drivers of non-universal behaviour in the case of the mapping latent shown in the middle panel of Fig.~\ref{fig:non-universality_in_dominant_latents}. The mapping latent  carries more significant information about non-universality at high $\sigma^{-1}$ (i.e. high mass). As discussed in Section~\ref{sec:latent_rep-universal_info}, this latent carries information about $\Omega_{\rm m}$, $\frac{\mathrm{d}\log \sigma}{\mathrm{d}\log M}$ and $N_{\mathrm{eff}}$. From the perspective of non-universality, we find that the information carried by $\Omega_{\rm m}$ is near-identical to that carried by the latent. To verify this explicitly, the dotted line shows that once $\Omega_{\rm m}$ is conditioned on (in addition to the other latents), the non-universality information in the mapping latent is close to zero.

The bottom panel shows the information about non-universality in the non-universal latent. As previously noted, this dominates the non-universal information, especially at low $\sigma^{-1}$ (i.e. low mass). 
Previously in Fig.~\ref{fig:cMI_latents_with_Dz}, we showed that the information in the non-universal latent is strongly related to the recent growth history after matter-dark-energy equality; the conditional MI between the latent and the growth function peaks at $z_{\rm late} \simeq 0.11$. We accordingly find that the growth factor at this particular redshift largely encodes the non-universal information content; the dashed line in Fig.~\ref{fig:non-universality_in_dominant_latents} traces a similar dependence on $\sigma^{-1}$. However, unlike for the other two latents, here we find there is a substantial difference between the information in this leading order dependence and the information in the latent itself.

\begin{figure}
	\includegraphics[width=\columnwidth]{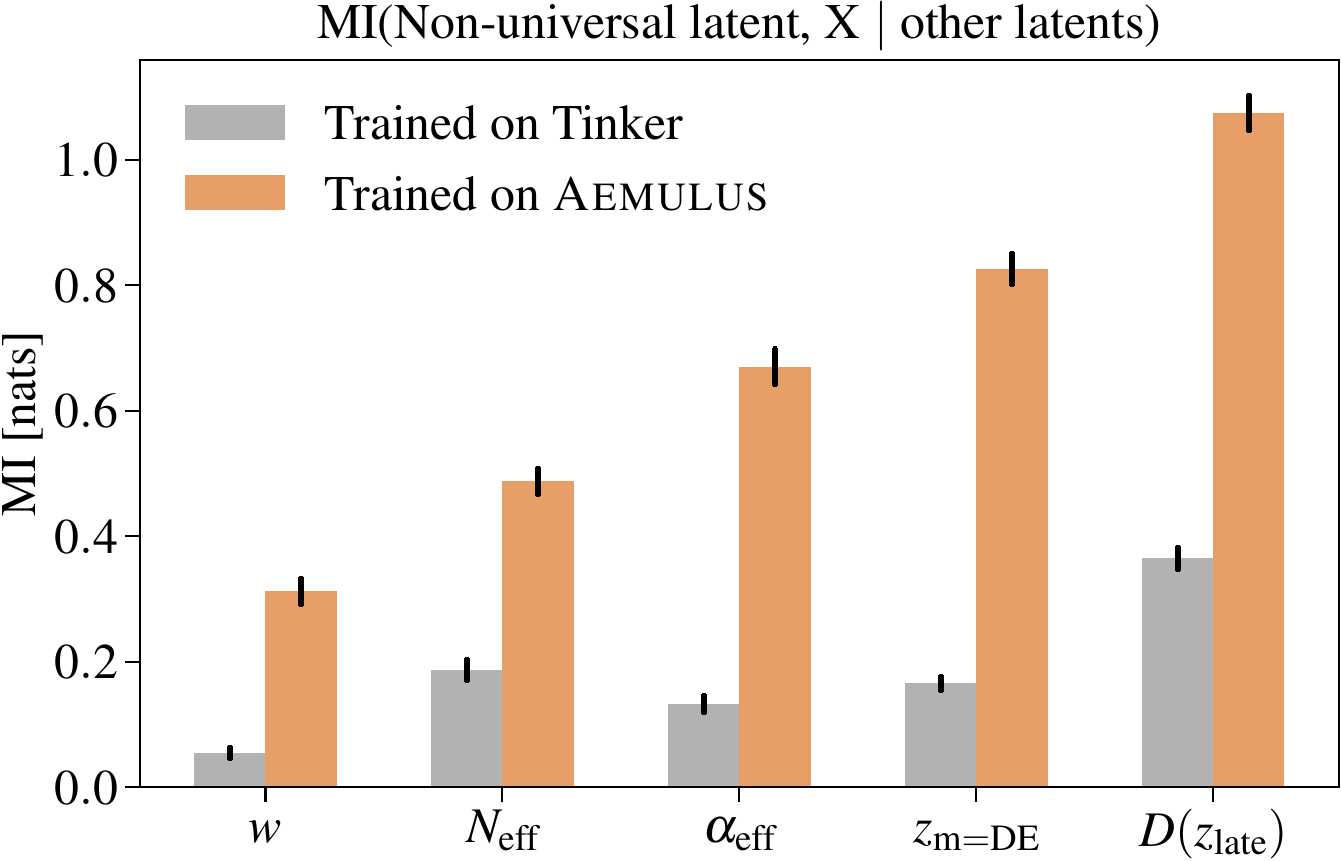}
	\caption{Conditional MI between the non-universal latent and X given the other two latent variables, where X is each of the cosmological quantities on the $x$-axis. We compare the MI of latents of two IVE models, separately trained on Tinker (grey) and \aemulus{} HMFs (orange); the former does not contain any non-universal information by construction. The non-universal latent has high information on parameters associated with the recent growth history, as well as $N_\mathrm{eff}$. The parameter $\alpha_\mathrm{eff}$ is the effective growth rate parametrization used by \citet{Ondaromallea2021}.}
	\label{fig:information_in_latentC-cMI_with_cosmo_parameters}
\end{figure}

To understand more fully what is driving the non-universality, Fig.~\ref{fig:information_in_latentC-cMI_with_cosmo_parameters} shows the MI between the non-universal latent and the most relevant cosmological parameter combinations (conditioned on the other two latents). 
We show this separately for two IVEs trained independently on Tinker HMFs, which are universal by definition, and \aemulus{} HMFs. This allows us to assess the amount of encoded non-universal information.
We find that the $D(z_{\rm late})$ encoding is a better description of the non-universal aspects compared to the $\alpha_{\rm eff}$ parametrization used by \citet{Ondaromallea2021}. The non-universal latent is also highly correlated with matter-dark-energy equality, but this does not maximize the information content as $D(z_{\rm late})$ does. By contrast, the Tinker latents contain only a small fraction of the information in the \aemulus{} latents about these quantities. The cosmological parameter dependence of $D(z_{\rm late})$ can be approximated using the results of \cite{Linder2005} as
\begin{equation}
D(z_{\rm late}) \simeq \left(\frac{1}{1+z_{\rm late}}\right)^{\Omega_{\rm m}(\bar{z})^{\gamma} } \,, \label{eq:Dz-param-dependence}
\end{equation}
where $\Omega_{\rm m}(\bar{z})$ is the matter density parameter at redshift $\bar{z} \simeq 0.05$, and $\gamma \simeq 0.55 + 0.02(1+w)$ for $w<-1$ and $\gamma \simeq 0.55 + 0.05(1+w)$ otherwise. The matter density parameter in turn  depends on its value at $z=0$, and $w$. Equation~\eqref{eq:Dz-param-dependence} is derived assuming that the growth rate is well-approximated by $\Omega_{\rm m}(z)^{\gamma}$ in $w$CDM. The derivation of equation~\eqref{eq:Dz-param-dependence} and a demonstration of its accuracy are given in Appendix \ref{sec:Dz-param}.

In addition to these growth-related indicators, we also find that the non-universal latent adds to the other latents significant information about the effective number of neutrino species, $N_{\mathrm{eff}}$. However, the added information on $N_{\mathrm{eff}}$ is subdominant to that about the growth history. Returning to the bottom panel of Fig.~\ref{fig:non-universality_in_dominant_latents}, the dotted line shows that once $N_{\mathrm{eff}}$ and $D(z_{\mathrm{late}})$ are conditioned upon, there is basically no information left about non-universality. 

\subsection{Connection between mass definition and non-universality}
\label{sec:mass_definition}

Our results apply to dark matter halo abundances defined using the $M_{\rm 200b}$ mass definition; this choice is motivated by the spherical collapse model and is commonly used (together with the 500\,$\rho_{\rm c}$ overdensity definition) in cluster cosmology analyses.
Previous work has shown that non-universality varies depending on the mass definition or equivalently, the mass distribution out to different physical scales \citep{Despali2016, Ondaromallea2021}; e.g. the 500$\rho_{\rm c}$ overdensity definition resolves only the inner parts of haloes, 200$\rho_{\rm b}$ the virial region, and the splashback mass the outer region \citep{Diemer2014, Adhikari2014, More2015, Diemer2020}. 

It has been shown that mass definitions probing the inner regions of haloes (500\,$\rho_{\rm c}$) tend to exhibit more non-universality than those that probe out to the outskirts of haloes (splashback) for cluster-sized haloes \citep{Diemer2020}. Our result that non-universality is tied to late-time growth can explain this finding from a physical perspective, since the inner regions of haloes are not as sensitive to late-time growth as the outer-regions. In this picture, we postulate that non-universality can originate because a given halo mass definition (in our case $M_{\rm 200b}$) artificially truncates cosmological information contained in the halo accretion history (in our case the late time growth). 
Conversely, in the case of a mass definition sensitive to the accretion history relating to the outskirts (and hence late time growth), that information would already be incorporated into the mass definition, meaning that the HMF would not need to be reparametrized to account for the extra accretion physics. We therefore expect a mass definition enclosing more of the density profile (and hence more information about accretion history) to lead to a more universal HMF.

This picture ties to our previous results in \citet{Lucie-Smith2023}, which showed that the universality in the density profiles of haloes may originate from a universality in the halo accretion histories themselves. Further, our interpretation implies that the larger amounts of non-universality seen in the HMF at higher redshifts \citep{Diemer2020} arise due to the fact that haloes at higher redshifts incorporate less of their accretion history within any given mass definition compared to $z=0$. This in turn may explain the close connection between the mass function across redshift and across cosmologies utilised by previous work including the cosmology-rescaling technique \citep{AnguloWhite2010}.

In addition to the effect of actual increase in mass through accretion, the mass of haloes can also appear to undergo a spurious `pseudo-evolution' simply due to redshift evolution of the background density used in the mass definition \citep{Diemer_2013}. We may therefore expect that this non-physical effect related to the overdensity definition links to $\Omega_{\rm m}$, and indeed we see that the non-universality encoded in the mapping latent is described by this parameter.

\section{Discussion}
\label{sec:discussion}
We have used deep learning to generate new insights into accurately modelling the HMF and understanding universal and non-universal information encoded within it. We trained a deep learning framework (IVE) to compress linear matter power spectrum (and, optionally, growth function) information into a low-dimensional latent space, that is then decoded into a halo mass function. The training set was generated from the existing \aemulus{} emulator \citep{McClintock2019}. We then interpreted the latent representation using MI to extract knowledge on the physical quantities required to model the HMF in \aemulus{}. We summarize our main findings below.

\begin{enumerate}

\item The linear growth function $D(z)$ does not carry information about the present-day HMF that is independent of $P(k, z=0)$ in the $w$CDM$+N_\mathrm{eff}$ parameter space. While $D(z)$ can play a physical role in halo growth, our architecture can extract relevant aspects of the growth history from $P(k, z=0)$. This is not possible when using $\sigma(R, z=0)$, since over a finite radius range the latter is a lossy compression of the former.

\item Three independent latent variables are necessary and sufficient to predict the HMF at $z=0$ to $\leq 0.25$~per~cent residuals for $M = 10^{13.2-15} \hmass$ in the  $w$CDM$+N_\mathrm{eff}$ parameter space. In other words, the IVE with three latents reproduces \aemulus{} essentially perfectly, since \aemulus{} itself produces halo mass functions with slightly better than 1 per cent accuracy. 

\item One latent variable primarily encodes universal information related to the mass variance $\sigma(M)$. It has a cosmological parameter dependence $\Omega_{\mathrm{m}}^{0.46} \sigma_8$ very similar to $S_8$. This can be predicted from the formation history and mass fluctuation variance of high-mass haloes.

\item The second latent variable encodes the factor $\Omega_{\rm m}^{0.35} \frac{\derv \log \sigma}{\derv \log M}$. This corresponds closely to  the `mapping' factor within  equation~\eqref{eq:halo_mass_function_for_MI} responsible for mapping from the multiplicity function $f(\sigma)$ to the HMF. The power of $\Omega_{\rm m}^{0.35}$ corresponds to evaluating the mapping at the formation time of low-mass haloes in our sample. The latent also carries some information about the effective neutrino number $N_\mathrm{eff}$. This latent carries information about non-universality (through its $\Omega_{\rm m}$ dependence) at the high-mass end of the HMF, i.e. $M\sim 10^{15} \, \mathrm{M_\odot}\, h^{-1}$.

\item The third latent variable primarily encodes non-universal information about the low-mass end of the HMF, i.e. $M\sim 10^{13} \, \mathrm{M_\odot}\, h^{-1}$. This non-universal information can be parametrized by the linear growth function after matter-dark-energy equality, specifically at $z_{\mathrm{late}} = 0.11$, with a contribution also from $N_{\mathrm{eff}}$.

\end{enumerate}

Recently, \citet{Ondaromallea2021} studied simulations designed to isolate the role of the linear growth rate and that of the local slope of the power spectrum. To do so they considered cosmologies with values of $\Omega_{\rm m}$ and $n_s$, respectively, that are well outside current data constraints. They demonstrated that a measure of growth rate in the recent past, which corresponds to the growth rate at $z \simeq 0.4$, is related to non-universality. They additionally suggested a smaller dependence of non-universality on the slope of the power spectrum; this dependence was later confirmed by \citet{Euclid2023HMF} using a similar methodology with extreme spectral tilts. 

Conversely, our conclusions on the origin of non-universality are drawn across the relevant cosmological parameter space for forthcoming surveys, without the need to construct extreme test cosmologies. We find a strong role for recent growth in establishing non-universality, but we find that growth since matter-dark-energy equality (and more specifically, at $z \simeq 0.1$) characterizes its role significantly better than the proxy suggested by \citet{Ondaromallea2021}. We do not find the local slope of the power spectrum or $n_s$ to play any significant role  in non-universality at $z=0$ within the currently viable cosmological parameter space we consider.

Overall, our deep learning technique has allowed us to segregate and quantify universal and non-universal aspects of information in the HMF, and to explain these aspects in terms of physical proxies related to cosmological halo formation. This approach builds on the finding in the existing literature that non-universality is related to growth history \citep{Courtin2010, Ondaromallea2021}; however, the existing literature is tied to the specific EPS-motivated functional form, whereas our more general approach allows us to gain physical insights that are challenging to obtain through empirical fitting formulae. 

In this paper, we  focused on training IVE models to predict the HMF at a single redshift $z=0$. Since the HMF exhibits strong non-universality with redshift evolution \citep{Reed2007, Tinker2008, Bhattacharya2011, Diemer2020} and forthcoming surveys will observe clusters out to $z= 2$ \citep{Adam2019, Cerardi2023}, future work will need to model both the cosmology and the redshift dependence of the HMF \citep[e.g.][]{McClintock2019, Bocquet2020, Artis2021}. The IVE framework we have presented in this paper can be straightforwardly extended to address this problem by adding redshift as an additional query. Our framework can also be applied to a range of other training data, including HMFs generated by other emulators such as \textsc{Dark Emulator} \citep{Nishimichi2019} and Mira-Titan \citep{Bocquet2020}, or directly to numerical simulations. Such generalizations would test how our findings extend beyond the domain of validity of \aemulus{}. Further, comparing the latent spaces of IVEs trained on HMFs using different halo definitions will allow us to explicitly test the connections between information in halo accretion histories and HMF non-universality discussed in Section~\ref{sec:mass_definition}.

The latent representation that IVE models learn can be applied to assist with the design of emulator training sets.  The training sets of existing emulators are generally designed to uniformly cover the cosmological parameter space \citep[e.g.][]{Heitmann2016, DeRose2019}. However, this approach does not take into account which directions within the space are more or less informative. 
By contrast, the three-dimensional latent space of the IVE corresponds to independent factors governing changes in the HMF. Therefore, by sampling this 3D latent space more uniformly, the accuracy of existing emulators could be improved using a small number of additional training simulations.  We propose that instead of adding simulations to uniformly sample the cosmological parameter space with higher density, simulations can be added such that the 3D latent space is more uniformly covered. Taken together with our proposed extensions to redshift dependence, alternative cosmological parameter spaces and halo definitions, the IVE-based approach thus promises new insights that will help tackle the crucial problem of predicting accurate theoretical HMFs for forthcoming data analyses.

\section*{Acknowledgements}

NG thanks Justin Alsing for helpful discussions on model training and Alessio Spurio Mancini for help with TensorFlow. LLS thanks Eiichiro Komatsu for useful discussions and for providing comments on the paper draft. We thank Raul Angulo, Benedikt Diemer and Lurdes Ondaro-Mallea for useful discussions. We also thank Thomas McClintock, Eduardo Rozo, Joseph DeRose, Jeremy Tinker and Risa Wechsler for their help with \aemulus{}. NG is supported by the UCL Graduate Research Scholarship and Overseas Research Scholarship, and the Perren Fund. This project has received funding from the European Research Council (ERC) under the European Union’s Horizon 2020 research and innovation programme (grant agreement No. 101018897 CosmicExplorer). HVP and LLS acknowledge the hospitality of the Aspen Center for Physics, which is supported by National Science Foundation grant PHY-1607611. The participation of HVP and LLS at the Aspen Center for Physics was supported by the Simons Foundation. The work of HVP was partially supported by the G{\"o}ran Gustafsson Foundation for Research in Natural Sciences and Medicine. AP has received funding from the European Union’s Horizon 2020 research and innovation programme under grant agreement No. 818085 GMGalaxies. DP is supported by the SNF Sinergia grant CRSII5-193826 “AstroSignals: A New Window on the Universe, with the New Generation of Large Radio-Astronomy Facilities''. We acknowledge support from the UCL Cosmoparticle Initiative,
including use of the Hypatia GPU cluster. We are grateful to Edd Edmondson for technical support.
\section*{Author Contributions}

\textbf{NG}: methodology, software, formal analysis, investigation, writing (first draft, review \& editing), visualization; \textbf{LL-S}: conceptualization, methodology, validation, writing (first draft, review \& editing), supervision; \textbf{HVP}: conceptualization, methodology, validation, writing (first draft, review \& editing), supervision, funding acquisition; \textbf{AP}: methodology, validation, writing (first draft, review \& editing), supervision, funding acquisition; \textbf{DP}: software, writing (review).\\

\section*{Data Availability}

The data underlying this article and the trained baseline IVE model are available in Zenodo at \url{https://doi.org/10.5281/zenodo.11278831}.


\bibliographystyle{mnras}
\bibliography{references}



\appendix

\section{Information content of the non-universal latent}
\label{sec:Dz-param}

In Section \ref{sec:latent_rep-nonuniversal_info} we showed that the leading-order contribution to the non-universal latent is given by the growth factor $D(z_{\mathrm{late}})$ evaluated at $z_{\mathrm{late}}\simeq 0.11$. We then gave an expression for the relation of $D(z_{\mathrm{late}})$ to cosmological parameters, which we now further explain.

\begin{figure}
	\includegraphics[width=0.95\columnwidth]{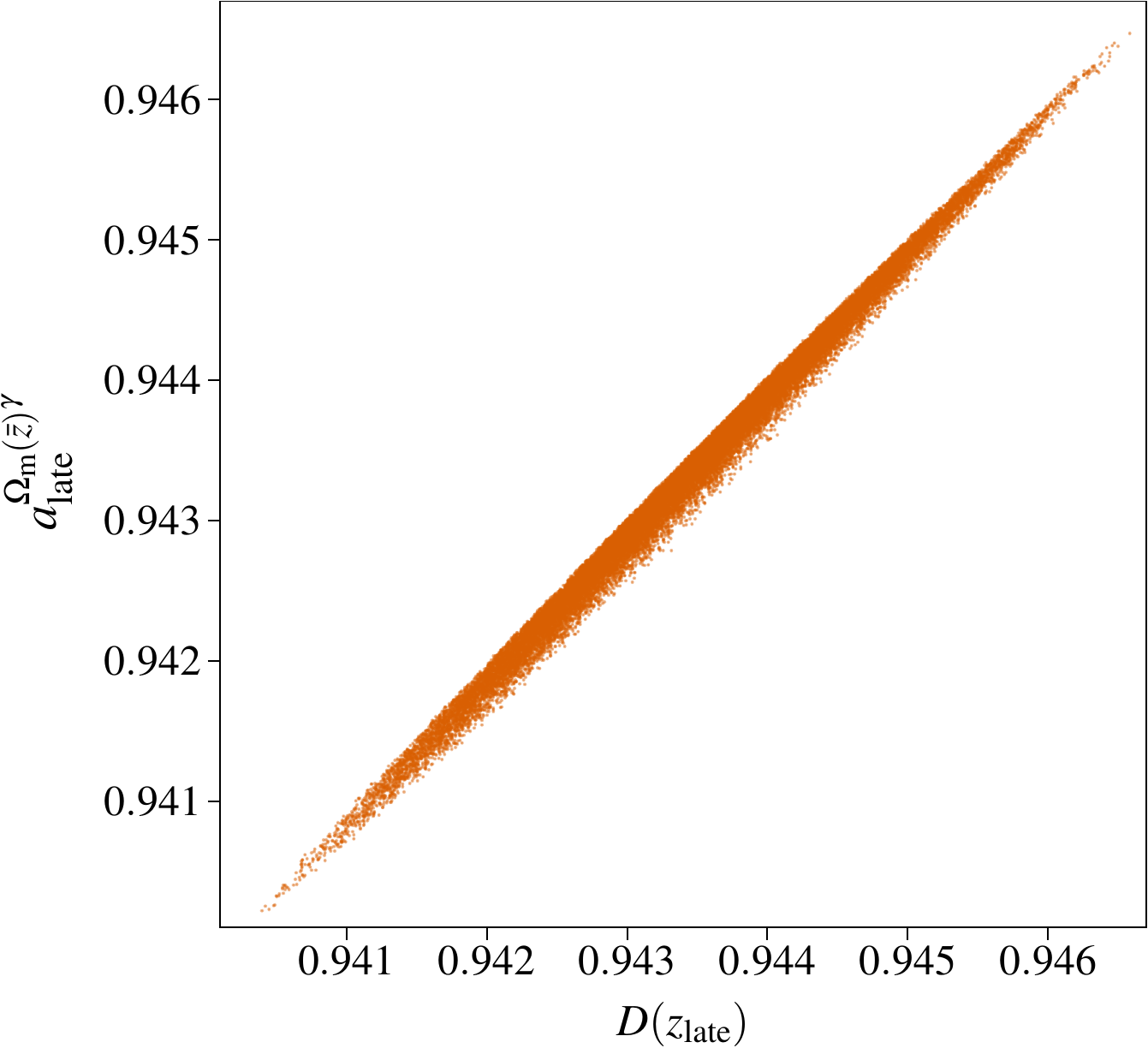}
	\caption{A demonstration that the approximate expression for $D(z_{\mathrm{late}})$ given in equation~\eqref{eq:Dzlate-appendix} closely approximates the numerically computed values of $D(z_{\mathrm{late}})$ across our parameter space.}
	\label{fig:Dz-approximation}
\end{figure}

\cite{Linder2005} provides an approximate expression for the linear growth function in wCDM cosmologies,
\begin{equation}
    g(a) \simeq \exp \left( \int_{-\infty}^{\ln a} \mathrm{d}\ln a' \left(\Omega_{\rm m}(z')^{\gamma} - 1 \right) \right)\,,
\end{equation}
where $z' = 1/a' - 1$ and $\gamma = 0.55 + 0.02 (1+w)$ for $w<-1$ or $0.55 + 0.05 (1+w)$ otherwise. We interpret our results in terms of the growth function normalized at the present day, $D(z) \equiv a g(a)/ g(1)$, and therefore
\begin{equation}
D(z) \simeq a\,\exp \left( \int_0^{\ln a} \mathrm{d}\ln a' \left(\Omega_{\rm m}(z')^{\gamma} - 1 \right) \right)\,.
\end{equation}
When evaluating this growth function at low redshift, $a \simeq 1$ and the range of the integral is small. Consequently, the integrand does not vary significantly and it may be factored out, giving the approximation
\begin{equation}
D(z_{\mathrm{late}}) \simeq a_{\rm late} \exp \left((\Omega_{\rm m}(\bar{z})^{\gamma} -1 ) \ln a_{\rm late} \right) = a_{\rm late}^{\Omega_{\rm m}(\bar{z})^{\gamma} }\,,\label{eq:Dzlate-appendix}
\end{equation}
where $\bar{z}$ is a representative redshift within the small integral range. There are a number of ways of choosing such a representative redshift, and in practice we find they all give similar results. The simplest is to define a representative scale factor,
\begin{equation}
\bar{a} \equiv \frac{\int_0^{\ln a_{\rm late}} \mathrm{d} \ln a' a' }{\int_0^{\ln a_{\rm late}} \mathrm{d} \ln a' } = \frac{a_{\rm late}-1}{\ln a_{\rm late}}\,.
\end{equation}
When we insert the value $z_{\mathrm{late}} = 0.11$, we obtain $\bar{a} \simeq 0.95$ and correspondingly $\bar{z} \simeq 0.05$, as quoted in the main text. For completeness we note that
\begin{equation}
\Omega_{\rm m}(\bar{z}) = \left(1+\frac{1- \Omega_{\rm m,0}}{\Omega_{\rm m,0}}\bar{a}^{-3 w}\right)^{-1}\,,
\end{equation}
making explicit the dependence on $\Omega_{\rm m,0}$ and $w$.

In Fig. \ref{fig:Dz-approximation} we show a scatter plot of the numerically integrated values of $D(z_{\mathrm{late}})$ against the approximate expression derived above and quoted in the main text, finding excellent agreement.

\section{Details of the method and implementation}

\subsection{Aemulus cosmological parameter space}
\label{appendix:aemulus_cosmological_parameter_space}
\begin{figure*}
	\includegraphics[width=\textwidth]{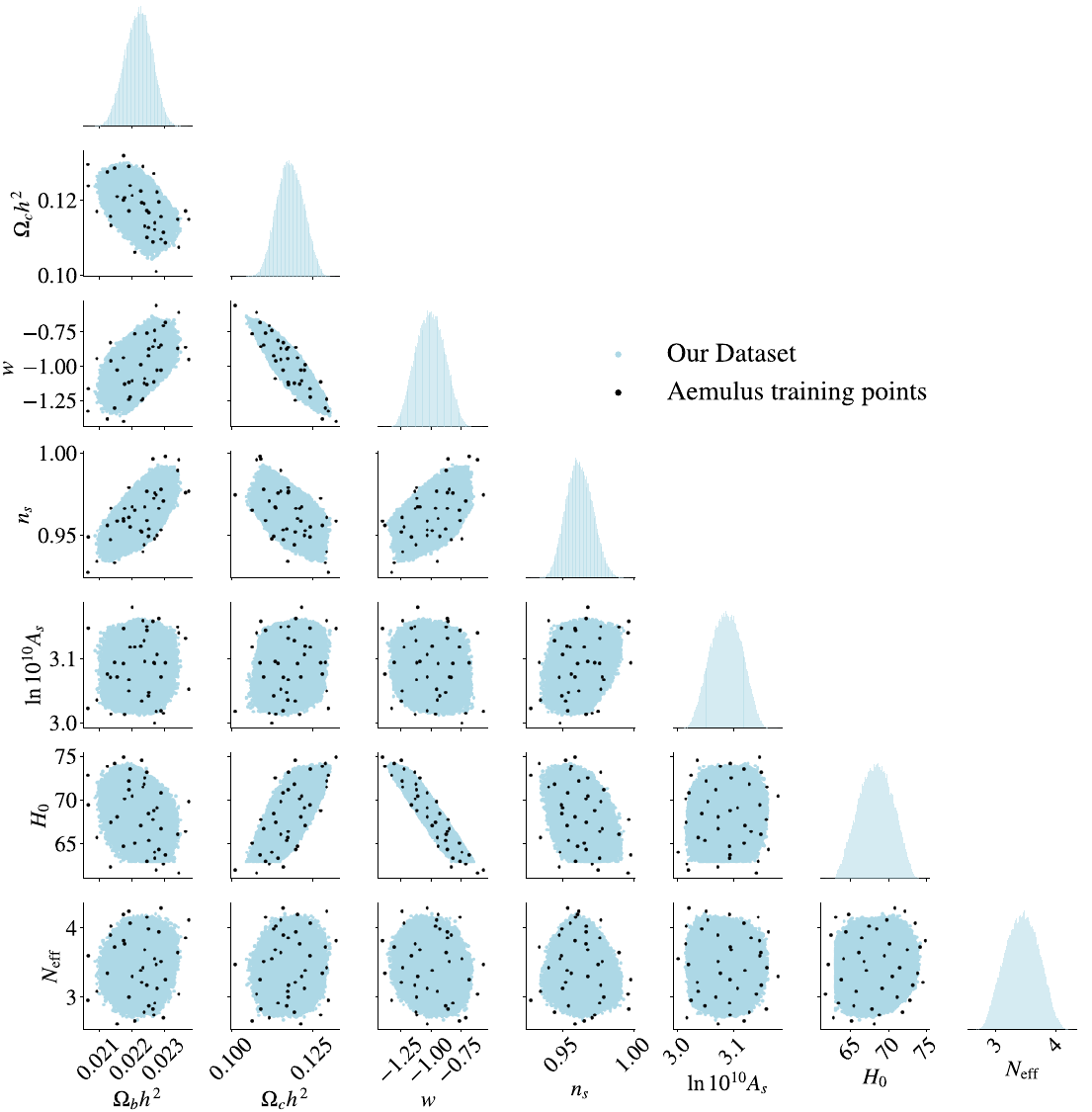}
	\caption{Cosmological parameter space covered by the dataset we use to train and test our IVE models (blue). Black points indicate the cosmologies used to train \aemulus{} \citep{McClintock2019}.}
	\label{fig:aemulus_cosmologies}
\end{figure*}

We describe here the procedure used to generate cosmological parameter samples used for our dataset so they lie within the domain of validity of \aemulus{} \citep{McClintock2019}. The cosmological parameter samples used to train \aemulus{} are drawn by projecting a unit Latin hypercube with 40 samples into parameter ranges spanning $\pm 4\sigma$ of CMB+BAO+SNIa constraints; these training cosmologies are shown in black in Fig.~\ref{fig:aemulus_cosmologies}. Our parameter samples are chosen using the same algorithm, but extending only to $\pm 3\sigma$; these samples are shown in blue. We choose $\pm 3\sigma$ rather than $\pm 4 \sigma$ because we sample the parameter space much more densely, so that allowing $\pm 4 \sigma$ would generate a large number of samples outside the region covered by \aemulus{}. 

The detailed steps to generate samples in Fig.~\ref{fig:aemulus_cosmologies} are:
\begin{enumerate}
	\item Concatenate two MCMC chains for the constraints which determine the \aemulus{} parameter space: Planck13+BAO+SNe (189623 samples) and \textit{e}WMAP9+BAO+SNe (72299 samples) for $w$CDM from \citet{Anderson2014}, where \textit{e}WMAP combines WMAP9 data with temperature power spectra from SPT and ACT.
	\item Obtain the eigenvectors and eigenvalues $\sigma_i^2$ of the concatenated chain by centring the data and performing principle component analysis; the eigenvalues give the variance in the direction of each eigenvector.
    \item Draw $10^7$ samples from a 7D unit Latin hypercube with sides $[0, 1]$.
	\item Project the Latin hypercube into the cosmological parameter space. Each dimension of the Latin hypercube is projected onto a single eigenvector, and the range for each dimension is scaled to $[-3, 3]\times \sigma_i$.  The MCMC chains vary only 6 cosmological parameters. The missing parameter, $N_\mathrm{eff}$, is obtained from  the 7th dimension of the hypercube, scaled appropriately to the minimum and maximum $N_{\mathrm{eff}}$ values in \textsc{Aemulus} training cosmologies. The \textsc{aemulus} training points were also obtained in this way.
	\item Discard all samples lying outside the convex hull of \textsc{Aemulus} training cosmologies. This eliminates the possibility for our network to have trouble learning the HMF due to incorrect ground truths produced by extrapolating the emulator outside its domain of validity.
\end{enumerate}

The procedure gives us a list of 628849 cosmological parameter samples, from which we randomly sample $10^5$ to produce our dataset. We verified that $10^5$ samples saturated the accuracy of the IVE.

As described in Section~\ref{sec:training_data}, \aemulus{} predicts the HMF by emulating fitting parameters to the multiplicity function $f(\sigma)$, and converts this into the HMF by calculating $\sigma(M)$ using CLASS \citep{Lesgourgues2011, Blas2011}. During our investigation, we have found that \aemulus{} misconfigured CLASS such that the power spectrum used to evaluate $\sigma(M)$ is always calculated using $w = -1$, even for $w \neq -1$ cosmologies. Because this misconfiguration was already present when training \aemulus{}, this does not affect the overall accuracy of \aemulus{}, but it does result in a correlation between the emulator's residuals and $w$. We find for $w > -1$, the emulator underpredicts the HMF especially at the high mass tail; the opposite is true for when $w < -1$. We expect this to have only a small effect on our results relating to non-universality, as this issue affects high mass haloes, whereas we find that non-universality affecting the high mass haloes is mostly related to $\Omega_{\rm m}$ rather than $w$. The recent growth history, on which $w$ has a stronger effect, affects mostly the non-universality for lower mass haloes where the residuals of \aemulus{} are small across the whole parameter space.

\subsection{IVE architecture}
\label{sec:IVE-architecture}
The IVE architecture consists of an encoder that compresses the input $P(k)$ (and, optionally, $D(z)$) into a low-dimensional latent representation, and a decoder which maps the latent representation and a given halo mass $M$ to the corresponding differential halo number density $\mathrm{d} n/\mathrm{d} \log M$.

The IVE takes as input a 1D array of either $P(k)$ at $z=0$, or a concatenation of $P(k)$ with $D(z)$. This input $\mathbf{x}$ is compressed into a low-dimensional latent representation by the encoder, which consists of a neural network with 6 fully-connected layers. Each fully-connected layer has 512 neurons, and each neuron follows $y=h(\mathbf{w} \cdot \mathbf{x} + \mathbf{b})$, where $\mathbf{w}$ and $\mathbf{b}$ are the weights and biases that are optimized when training the network, and $h$ is the activation function. We use leaky ReLU activation \citep{Maas2013} for all hidden layers except the last layer, for which we use a tanh activation to ensure smoothness of the encoder output. Weights are initialized with the \citet{He2015} uniform initialization scheme  for layers with leaky ReLU activation, and with \citet{Glorot2010} uniform initialization for the layer with tanh activation.

The encoder maps from the input $\mathbf{x}$ to a multivariate distribution in the latent space $p(\boldsymbol{\kappa}|\mathbf{x})$, where $\boldsymbol{\kappa}$ denotes the latent representation. (We use $\boldsymbol{\kappa}$ instead of the conventional $\mathbf{z}$ to avoid possible confusion with redshift.) The encoder approximates the latent distribution by a factorized Gaussian, $p(\boldsymbol{\kappa}|\mathbf{x}) = \prod_i^K \mathcal{N}(\mu_i(\mathbf{x}), \sigma_i(\mathbf{x}))$, where $K$ is the dimensionality of the latent space and $\mu_i$, $\sigma_i$ are the means and standard deviations of the Gaussian distribution for each component of the latent representation. Using this assumption, the encoder maps the inputs to $\mu_i$ and $\sigma_i$.

The decoder first samples a latent vector $\boldsymbol{\kappa}$ from the encoded Gaussian distribution. This, together with a query $\log M$, is passed into another neural network with 6 fully-connected layers, with the same number of neurons per layer, the same activation functions, and the same initialization schemes as the encoder. The output from the decoder is an estimate of $\log\, \mathrm{d} n/\mathrm{d} \log M$ at the given halo mass. 

\subsection{The regularization term in the loss function}
\label{app:regloss}
The IVE loss function was discussed in Section~\ref{sec:loss-function}. Here we consider further the second term in  equation~\eqref{eq:loss_function}, which is the Kullback-Leibler (KL) divergence \citep{Kullback1951}, measuring how close the latent distribution returned by the encoder $p(\boldsymbol{\kappa}|\mathbf{x})$ is to a prior distribution $q(\boldsymbol{\kappa})$. We choose as the prior distribution a diagonal standard Gaussian $q(\boldsymbol{\kappa}) = \prod_i^K \mathcal{N}(\mu_i=0, \sigma_i=1)$. This leads to a closed form for the regularization term:
\begin{equation}\label{eq:loss_function-KL_term}
	\mathcal{D}_{KL}(p(\boldsymbol{\kappa}|\mathbf{x}) \| q(\boldsymbol{\kappa})) = -\frac{1}{2}\sum_{i}^K \left(1 +  \log \left(\sigma_{i}^{2}\right)-\mu_{i}^{2}-\sigma_{i}^{2}\right) \,.
\end{equation}
The diagonal standard Gaussian prior encourages an interpretable latent space: it ensures continuity and completeness in the latent space since all samples are mapped to a localized region in the latent space \citep[see][]{Burgess2018}, and it promotes independence between latent variables since the marginal latent distribution $p(\boldsymbol{\kappa}) = \int p(\boldsymbol{\kappa}|\mathbf{x}) p(\mathbf{x})\derv \mathbf{x}$ is encouraged to be diagonal. This causes each latent variable to learn an independent factor of variation governing the HMF, which is crucial for the latent representation to be interpretable.

\subsection{Training procedure}
\label{app:annealing}

In Section \ref{sec:IVE-optimisation} we outlined our use of a $\beta$-annealing scheme to improve the prediction accuracy while maintaining a disentangled latent space. Here, we give more details on this procedure. We used a training data batch size of 200 and a learning rate of $5\times 10^{-4}$, and varied $\beta$ as a function of epoch $t$ according to a generalized logistic function of the form
\begin{equation}
	\beta(t) = \beta_i + \frac{\beta_f - \beta_i}{1 + \exp({-r(t-t_{\rm c})})} \,,
\end{equation}
where $\beta_i$ and $\beta_f$ are the initial and final values that $\beta$ asymptotes to, $r$ controls the rate of decrease, and $t_{\rm c}$ is the epoch where the rate of decrease in $\beta$ is maximum. 

We performed experiments with different $\beta_i$, $\beta_f$, $r$, and $t_{\rm c}$, and chose as our final models those with (i) a prediction accuracy that is comparable to when the model is trained with $\beta=0$, and (ii) the lowest possible MI between latent variables. In particular, we ensured that the MI between latents remained below the MI between the non-universal latent and the ground truth HMF. 

To achieve this, we performed a three-stage $\beta$-annealing procedure. In the first stage, we set $\beta_i = 3.6\times 10^{-5}$, $\beta_f = 10^{-8}$, $r=0.001$ and $t_{\rm c}=2000$. 
With this set of parameters, we were able to achieve an accuracy comparable to that of a three-latent IVE model trained with $\beta=0$, but at the expense of insufficient disentanglement. In particular, the MI between the non-universal latent and the other latents was higher than that between the former and the ground truth HMF. We performed a second and third stage of annealing to improve both the model accuracy and the latents' disentanglement. In the second stage, we (i) resumed the training starting from epoch 5000, where the accuracy matched that of a two-latent IVE model and the MI between all latents was $< \mathcal{O}(10^{-4})$~nats, and (ii) used $\beta_i=1.4\times 10^{-6}$, $\beta_f = 10^{-7}$, $r=0.0015$ and $t_{\rm c}=6000$. This set of parameters ensured that $\beta(t)$ decreased at a slower rate than in the first stage, which turned out to be crucial to maintain the disentanglement. In the third stage, we (i) resumed training at epoch 6000, where the prediction accuracy was again comparable to a two-latent IVE but the MI between the latents was $\sim \mathcal{O}(10^{-4})$~nats, and (ii) used $\beta_i = 8.485\times 10^{-7}$, $\beta_f = 10^{-7}$, $r=0.002$ and $t_{\rm c}=9000$. We trained the model for a further 6250 epochs until $\beta$ plateaued to its $\beta_f$ value. This three-stage process allowed us to finally achieve a sufficient level of disentanglement while reaching high prediction accuracy.

Once the $\beta$-annealing procedure finished, we further refined the prediction accuracy of the model to ensure it is comparable to the same model trained with $\beta=0$. We increased the batch size from 200 to 500 and halved the learning rate, and continued to double the batch size and halve the learning rate every time the loss on the validation set ceased to improve over 40 training epochs, until a maximum batch size of 8000 (limited by GPU memory). This last training phase improved the accuracy without affecting the level of disentanglement; $\beta$ was kept constant at $10^{-7}$ during this stage. Decreasing the learning rate reduces the possibility that the optimizer will overstep the minimum \citep{Alsing2020}, while increasing the batch size increases the accuracy of gradient direction estimates. Our models reached convergence in 311 additional epochs when the validation loss saturated. In total, training the IVE model (including all $\beta$-annealing and prediction accuracy refinement) lasted 12561 epochs.


\bsp	
\label{lastpage}
\end{document}